\documentclass[12pt, draftclsnofoot, onecolumn]{IEEEtran}
\usepackage{amssymb}
\usepackage{amsmath}
\usepackage{cite}
\usepackage{url}
\usepackage{xcolor}
\usepackage{cite,graphicx,amsmath,amssymb}
\usepackage{subfigure}
\usepackage{citesort}
\usepackage{fancyhdr}
\usepackage{mdwmath}
\usepackage{mdwtab}
\usepackage{caption}
\usepackage{amsthm}
\usepackage{setspace}
\usepackage{multirow}
 \usepackage{epstopdf}
\newtheorem{remark}{Remark}
\newtheorem{theorem}{Theorem}

\newtheorem{lemma}{Lemma}

\newtheorem{corollary}{Corollary}

\usepackage{algorithm}
\usepackage{algorithmic}

\begin{document}

\title{Resource Allocation in Intelligent Reflecting Surface Assisted NOMA Systems}
\author{
Jiakuo~Zuo,
Yuanwei~Liu,~\IEEEmembership{Senior Member,~IEEE,}
        Zhijin~Qin,~\IEEEmembership{Member,~IEEE,}
        and Naofal Al-Dhahir,~\IEEEmembership{Fellow,~IEEE}
\thanks{Part of this work has been submitted to the IEEE International Conference on Communications Workshop on NOMA for 5G
and Beyond, Dublin, Ireland, June 7-11, 2020.~\cite{Zuo2020}}
\thanks{J. Zuo is with Nanjing University of Posts and Telecommunications,
Nanjing, China. (email:zuojiakuo@njupt.edu.cn).}
\thanks{Y. Liu and Z. Qin are with Queen Mary University of London, London,
UK. (email:\{yuanwei.liu, z.qin\}@qmul.ac.uk).}
\thanks{N. Al-Dhahir is with the Department of Electrical and Computer Engineering, The University of Texas at Dallas, Richardson,
TX 75080 USA. (email:aldhahir@utdallas.edu).}
 }
\maketitle
\begin{abstract}
This paper investigates the downlink communications of intelligent reflecting surface (IRS) assisted non-orthogonal multiple access (NOMA) systems. To maximize the system throughput, we formulate a joint optimization problem over the channel assignment, decoding order of NOMA users, power allocation, and reflection coefficients. The formulated problem is proved to be NP-hard. To tackle this problem, a three-step novel resource allocation algorithm is proposed. Firstly, the channel assignment problem is solved by a many-to-one matching algorithm. Secondly, by considering the IRS reflection coefficients design, a low-complexity decoding order optimization algorithm is proposed. Thirdly, given a channel assignment and decoding order, a joint optimization algorithm is proposed for solving the joint power allocation and reflection coefficient design problem. Numerical results illustrate that: i) with the aid of IRS, the proposed IRS-NOMA system outperforms the conventional NOMA system without the IRS in terms of system throughput; ii) the proposed IRS-NOMA system achieves higher system throughput than the IRS assisted orthogonal multiple access (IRS-OMA) systems; iii) simulation results show that the performance gains of the IRS-NOMA and the IRS-OMA systems can be enhanced via carefully choosing the location of the IRS.
\end{abstract}


\section{Introduction}
 Recently, non-orthogonal multiple access (NOMA) has received considerable attention for its great potential to support massive connectivity and enhance spectrum efficiency. It has been included in the next generation digital television standard and the 3rd Generation Partnership Project (3GPP) Long Term Evolution-Advanced (LTE-A) standard~\cite{liu2018non,3GPPStudy}. Different from the conventional orthogonal  multiple access (OMA), NOMA allows multiple users to access the same orthogonal resource block, such as frequency band, time slot, and spatial direction. The successive interference cancellation (SIC) technique is employed at the receiver side. Particularly, the users with better channel conditions are capable of removing the intra-channel interference from the users with poor channel conditions. The performance gains brought by NOMA have been investigated under various scenarios~\cite{dai2018survey,qin2018user,ding2017application,liu2017non}.

The intelligent reflecting surface (IRS) is one of the promising solutions to improve the network coverage in future wireless networks~\cite{gong2019towards,zhao2019survey}. IRS comprises a large number of passive elements and each element can independently reflect the incident signal by adjusting the reflection coefficients, including phase shift and amplitude, so that the received signal power can be boosted at the receiver. Different from the traditional amplify-and-forward relay, the IRS does not have the signal processing capability. Instead, it only reflects the signals in a passive way, which makes it more energy-efficient. Moreover, the IRS is also different from the active intelligent surface based massive multiple-input multiple-output (MIMO), which suffers from the high hardware cost and power consumption. IRS can yield superior performance by increasing the number of elements with affordable hardware cost and tolerable power consumption~\cite{tang2019wireless,jung2018performance}.

With the ability to control the channel conditions by adjusting the phase shift and amplitude of the IRS elements, it is of great interest to investigate the potential benefits of IRS assisted NOMA (IRS-NOMA) systems by utilizing the IRS to provide additional paths to construct a stronger combined channel gain.

\subsection{Related Works}
\subsubsection{Resource Allocation for NOMA Systems}
In this paper, we focus on the channel assignment and power allocation problem for NOMA systems. Typically, the joint channel assignment and power allocation optimization problem is of mixed integer type, which is non-convex and difficult to solve directly. There are mainly two approaches to solve such type of problems in the literature, including the suboptimal approach~\cite{di2016sub,zhao2017spectrum,liu2018super,shi2019energy} and the optimal approach~\cite{fang2018joint,sun2017optimal,sun2018robust}.

Specifically, for the NOMA systems in~\cite{di2016sub}, the channel assignment problem was first solved by a many-to-many matching algorithm and then the power allocation problem was solved by geometric programming. In~\cite{zhao2017spectrum}, a swap-operation enabled matching algorithm was proposed for solving the channel assignment problem in NOMA enhanced heterogeneous networks (HetNets) and sequential convex programming was adopted to update the power allocation. The channel assignment problem in energy-efficient NOMA systems~\cite{liu2018super} was modeled as a super-modular game and a greedy bidirectional channel matching algorithm was proposed. For a given channel matching, the non-convex power allocation problem was transformed to a convex problem by the successive convex approximation (SCA) method. In~\cite{shi2019energy}, by decoupling the channel and power variables, the channel assignment problem was solved by the exhaustive search method and the power allocation problem was solved via the SCA method. In~\cite{fang2018joint}, the channel assignment and the power allocation problems were solved jointly. By relaxing the binary channel indicator variables into continuous variables, the relaxed channel assignment and power allocation problem was solved by the Lagrangian approach, where optimal closed-form power allocation expressions were derived. Moreover, in~\cite{sun2017optimal,sun2018robust}, resource allocation was formulated as a monotonic optimization problem, and an optimal joint channel assignment and power allocation algorithm was proposed.

\subsubsection{Reflection Coefficient Design for IRS Assisted Wireless Systems}
There have been extensive works on IRS assisted wireless communication systems, such as the IRS assisted MIMO~\cite{ning2019intelligent,zhang2019capacity,chu2019intelligent}, IRS assisted  orthogonal frequency division
multiplexing (OFDM)~\cite{yang2019intelligent}, IRS assisted unmanned aerial vehicle (UAV) systems~\cite{li2019reconfigurable}, IRS assisted simultaneous wireless information and power transfer (SWIPT) systems~\cite{wu2019joint}, and IRS-NOMA systems~\cite{yang2019intelligentNOMA,ding2019simple,li2019joint,mu2019exploiting,zhu2019power,fu2019reconfigurable}. Particularly, a joint power allocation and phase shifts optimization problem was first studied for the IRS-NOMA systems in~\cite{yang2019intelligentNOMA}. The formulated problem was solved based on the alternating optimization algorithm and semidefinite relaxation (SDR). A new decoding order searching algorithm was proposed by maximizing the combined channel gain of each user. In~\cite{ding2019simple}, a multiple-input single-output (MISO) IRS-NOMA transmission model with fixed decoding order was considered. Assuming ideal beamforming, the phase shift was optimized by maximizing the signal-interference-plus-noise (SINR) with zero-forcing beamforming. Under the finite resolution beamforming assumption, by applying on-off control, a low-cost implementation structure was proposed to control phase shifts. In~\cite{li2019joint}, by assuming the perfect SIC decoding order, an effective second-order cone programming (SOCP)-alternating direction method of multipliers (ADMM) based algorithm was proposed for MISO IRS-NOMA system. To reduce the complexity, a zero-forcing based suboptimal algorithm was also introduced. In~\cite{mu2019exploiting}, two cases of reflection coefficients design for MISO IRS-NOMA systems were considered. For the ideal IRS scenario, both phase shifts and amplitudes were optimized. For the non-ideal IRS scenario, the amplitudes were fixed and only the phase shifts were optimized. For  both cases, the optimal decoding order was obtained by exhaustive search. In~\cite{zhu2019power}, the power efficient MISO IRS-NOMA system under quasi-degraded channels was studied. To ensure that the system achieves the capacity region with high probability, an improved quasi-degradation condition was proposed by using IRS. Moreover, the beamforming vectors and IRS phase shift matrix were optimized jointly based on the alternating optimization algorithm and the SDR method. In~\cite{fu2019reconfigurable}, an alternating difference-of-convex (DC) algorithm was proposed to solve the joint beamforming and phase shifts optimization problem. Furthermore, a low-complexity user ordering scheme was proposed by considering the phase shifts and target data rates.

\subsection{Motivation and Challenges}
The network coverage can be improved by introducing the IRS. However, the formulated optimization problems become non-trivial to solve, since the reflection coefficients of IRS are usually coupled with the other variables, such as transmit power and beamforming vector. Therefore, efficient algorithms should be carefully designed for IRS assisted wireless communication systems.

To our best knowledge, there is no existing work on joint optimization of the channel assignment, decoding order, power allocation, and reflection coefficients for IRS-NOMA systems. For the resource allocation in the IRS-NOMA systems, we identify the major challenges as follows:
\begin{itemize}
  \item The joint optimization problem is NP-hard, which makes the formulated resource allocation problem non-trivial to solve.
  \item In the IRS-NOMA systems, the SIC decoding order depends on the combined channel from both the direct link and the reflection link, which are determined by the IRS reflection coefficients as well. Therefore, determining the optimal decoding order for NOMA users is challenging.
  \item The transmit power and reflection coefficients are highly coupled.
\end{itemize}
\subsection{Contributions}
The main contributions of this paper are summarized as follows:
\begin{enumerate}
  \item We propose a multi-channel downlink communications IRS-NOMA framework, where multiple users are allowed to be flexibly assigned to the same channel. We formulate the system throughput maximization problem subject to SIC decoding conditions and IRS reflection coefficients constraints by jointly optimizing the channel assignment, decoding order, power allocation, and reflection coefficients.
   \item To solve the formulated problem, we decompose the original problem into three sub-problems. Firstly, a low complexity many-to-one matching algorithm is proposed for the channel assignment. Secondly, a low-complexity SIC decoding order optimization algorithm is proposed by maximizing the overall combined channel gains. Thirdly, we propose an efficient algorithm by invoking the alternating optimization approach to optimize the power allocation and reflection coefficients alternately.

  \item We demonstrate that the proposed channel assignment and low-complexity decoding order optimization algorithm can achieve near-optimal performance. The proposed three-step resource allocation algorithm for the IRS-NOMA system can improve the system throughput. Moreover, we will demonstrate that the system performance can be enhanced by deploying the IRS near the receivers.

\end{enumerate}
\subsection{Organization}
The rest of this paper is organized as follows. In Section II, the system model is introduced and the formulated resource allocation problem is presented. In Section III, we propose efficient algorithms to solve the resource allocation problem for the IRS-NOMA systems. Numerical results are presented in Section IV, which is followed by the conclusions in Section V.

Notations: $\mathbb{C}^{M \times 1}$ denotes a complex vector of size \emph{M}, diag(\textbf{x}) denotes a diagonal matrix whose diagonal elements are the corresponding elements in vector \textbf{x}. ${\textbf{x}}^{H}$ denotes the conjugate transpose of vector \textbf{x}. The notations Tr(\textbf{X}) and rank(\textbf{X}) denote the trace and rank of matrix \textbf{X}, respectively, while $\angle x$ denotes the phase of a complex number \emph{x}. The functions real(\emph{x}) and imag(\emph{x}) denote the real and imaginary part of a complex number \emph{x}.
\section{System Model And Problem Formulation}
\subsection{System Model}
\begin{figure}[!t]
\centering
\includegraphics[scale=0.6]{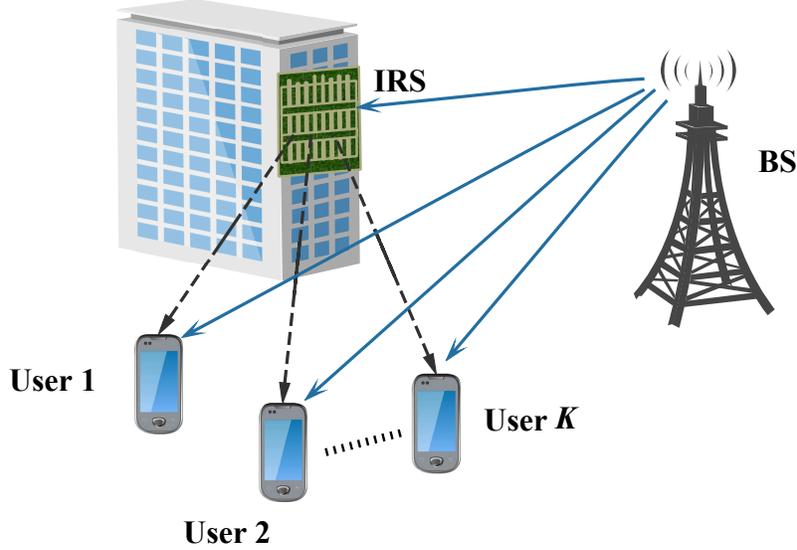}
\caption{Illustration of the downlink IRS-NOMA system.}
\label{system model}
\end{figure}

As shown in Fig. \ref{system model}, we consider the downlink transmissions in a IRS-NOMA system where there is one base station (BS), one IRS and \emph{K} users. Let ${\cal K} = \left\{ {1,2, \cdots ,K} \right\}$  denote the set of users. The total bandwidth $B$ is equally divided into a set of channels, denoted by ${\cal N} = \left\{ {1,2, \cdots ,N} \right\}$, each with the bandwidth of $W = {B \mathord{\left/ {\vphantom {B N}} \right. \kern-\nulldelimiterspace} N}$. Let ${{\cal K}_n}$ denote the set of the users assigned to the \emph{n}-th channel and ${K_n} = \left| {{{\cal K}_n}} \right|$ denote the maximum number of users assigned to the \emph{n}-th channel, $n \in {\cal N}$. Assume that each user is assigned only one channel and each channel can be assigned to ${K_n}$ users at most, then we have ${{\cal K}_n} \cap {{\cal K}_{\overline n }} = \emptyset$ and ${ \cup _{n \in {\cal N}}}{{\cal K}_n} = {\cal K}$, where $n \ne \overline n$. The IRS-NOMA system becomes an IRS-OMA system when ${K_n}{\rm{ = }}1$. The IRS consists of $M$ passive reflecting elements, denoted by ${\cal M} = \left\{ {1,2, \cdots ,M} \right\}$. Let $\boldsymbol{\Theta} = {\rm{diag}}\left\{ {{\lambda _1}{e^{j{\theta _1}}},{\lambda _2}{e^{j{\theta _2}}}, \cdots ,{\lambda _M}{e^{j{\theta _M}}}} \right\}$ denote the reflection coefficients matrix of the IRS, where ${\theta _m} \in \left[ {0,2\pi } \right]$ and ${\lambda _m} \in \left[ {0,1} \right]$ denote the phase shift and amplitude of the \emph{m}-th reflecting element, respectively~\cite{yang2019intelligent}.

The superposition symbol ${x_n}$ to be transmitted on the $n$-th channel is
\begin{align}\label{signal x n}
{x_n} = \sum\limits_{k = 1}^K {{\delta _{n,k}}\sqrt {{p_{n,k}}} {s_{n,k}}},
\end{align}
where ${p_{n,k}}$ is the power allocated to the $n$-th channel used by user $k$, ${s_{n,k}}$ is the symbol transmitted by user $k$ over the $n$-th channel, and ${\delta_{n,k}} \in \left\{{0,1} \right\}$ indicates whether the $n$-th channel is assigned to user $k$.

The signal received at user $k$ over the $n$-th channel is
\begin{align}\label{signal_y_n_k}
{y_{n,k}} = \left( {\textbf{g}_{n,k}^H\boldsymbol{\Theta}{\textbf{f}_n} + {h_{n,k}}} \right){x_n} + {z_{n,k}},
\end{align}
where ${\textbf{g}_{n,k}} \in \mathbb{C}^{M \times 1}$ is the channel gain for the $n$-th channel between the IRS and user $k$ , ${\textbf{f}_{n}} \in \mathbb{C}^{M \times 1}$ is the channel gain for the $n$-th channel between the BS and the IRS, and ${h_{n,k}}$ is the channel gain for the $n$-th channel between the BS and user $k$ .

 Without loss of generality, we assume that all the BS-User link channels $\{{h_{n,k}}\}$ and IRS-User link channels $\{{\textbf{g}_{n,k}}\}$ are mutually independent and follow Rayleigh fading, $n \in {\cal N}, \ k \in {\cal K}$. Rician fading channel model is adopted for the BS-IRS link channels $\{{\textbf{f}_{n}}\}$ ($n \in {\cal N}$), which is modeled as
\begin{align}\label{BS-IRS channel}
\textbf{f}_n = \sqrt {\frac{\kappa }{{1{\rm{ + }}\kappa }}{\rm{  }}} {{\textbf{f}}_n^{{\rm{LoS}}}} + \sqrt {\frac{1}{{1{\rm{ + }}\kappa }}{\rm{  }}} {{\textbf{f}}_n^{{\rm{NLoS}}}}
 \end{align}
where $\kappa$ is the Rician factor,  ${\textbf{f}_n^{\rm Los}}$ and ${\textbf{f}_n^{\rm NLoS}}$ are the line-of-sight (LoS) component and non-LoS (NLoS) component, respectively. The elements of ${\textbf{f}_n^{\rm NLoS}}$ are assumed to be independent and follow the Rayleigh fading model.

The SIC decoding order is an essential issue for NOMA systems, where the optimal decoding order is determined by the channel gains. However, in IRS-NOMA systems, the combined channel gains can be modified by tuning the IRS reflection coefficients. Denote ${\pi _n}\left( k \right)$ as the decoding order for user $k$ transmitting over the $n$-th channel. Then, ${\pi _n}\left( k \right) = j$ means that user $k$ is the $j$-th signal to be decoded at the receiver. The achievable capacity for user $k$ on the $n$-th channel can be expressed as
\begin{align}\label{Rate_n_k}
{R_{n,k}} =  {\log _2} \left( {1 + \frac{{{\delta _{n,k}}{p_{n,k}}{{\left| {\textbf{g}_{n,k}^H\boldsymbol{\Theta}{\textbf{f}_n} + {h_{n,k}}} \right|}^2}}}{{{{\left| {\textbf{g}_{n,k}^H\boldsymbol{\Theta}{\textbf{\textbf{f}}_n} + {h_{n,k}}} \right|}^2}{P_{n,k}}  + {\sigma ^2}}}} \right),
\end{align}
where ${P_{n,k}} = \sum\limits_{{\pi _n}\left( i \right) > {\pi _n}\left( k \right)} {{\delta _{n,i}}{p_{n,i}}}$.

Assume that ${\pi _n}\left( k \right) \le {\pi _n}\left( {\overline k } \right) $, then the capacity when user $\overline k $ decodes user $k$'s signal is
\begin{align}\label{Delta_n_k}
{R_{n,\bar k \to k}} = {\log _2}\left( {1 + \frac{{{\delta _{n,k}}{p_{n,k}}{{\left| {{\bf{g}}_{n,\bar k}^H{\bf{\Theta }}{{\bf{f}}_n} + {h_{n,\bar k}}} \right|}^2}}}{{{{\left| {{\bf{g}}_{n,\bar k}^H{\bf{\Theta }}{{\bf{f}}_n} + {h_{n,\bar k}}} \right|}^2}{P_{n,k}} + {\sigma ^2}}}} \right),
\end{align}

To guarantee that user $\overline k$ can decode the information of user \emph{k} successfully under the decoding order ${\pi _n}\left( {\overline k } \right) \ge {\pi _n}\left( k \right)$, the SIC decoding condition ${R_{n,\overline k  \to k}} \ge {R_{n,k}}$ should be guaranteed~\cite{liu2016fairness,cui2017optimal}. For example, assume that there are three users accessing the $n$-th channel and the SIC decoding order is ${\pi _n}\left( k \right){\rm{ = }}k,\ k = 1,2,3$. Then, the SIC decoding conditions at user 2 and user 3 should satisfy the following condition: ${R_{n,2 \to 1}} \ge {R_{n,1}},{R_{n,3 \to 1}} \ge {R_{n,1}},{R_{n,3 \to 2}} \ge {R_{n,2}}$, $n \in {\cal N}$.
\subsection{Problem Formulation for the IRS-NOMA System}
To maximize the system throughput, we should jointly optimize the channel allocation, decoding order, power allocation and reflection coefficients. The optimized problem can be formulated as
\begin{subequations}
\begin{align}\label{original problem}
  & {\rm (\textbf{P1})} \ \ \mathop {\max }\limits_{\boldsymbol{\delta} ,\boldsymbol{\pi} ,\textbf{p},\boldsymbol{\Theta} } \sum\limits_{n = 1}^N {\sum\limits_{k = 1}^K {{R_{n,k}}} }, \\
  &{\rm s.t.} \ {R_{n,\overline k  \to k}} \ge {R_{n,k}},\ {\rm if} \ {\pi _n}\left( k \right) \le {\pi _n}\left( {\overline k } \right), \ n \in {\cal N}, \ k, \ \overline k  \in {\cal K}, \label{original Constraints b}\\
  &  {\ \ \ \ \ R_{n,k}} \ge R_{\min }, \ n \in {\cal N}, \ k \in {\cal K}, \label{original Constraints c}\\
  & \ \ \ \ \ \sum\limits_{n = 1}^N {\sum\limits_{k = 1}^K {{\delta _{n,k}}{p_{n,k}}} }  \le {P_{\max }}, \label{original Constraints d}\\
 &  \ \ \ \ \ \left| {{\boldsymbol{\Theta} _{m,m}}} \right| \le 1, \ m \in {\cal M}, \label{original Constraints e}\\
 &  \ \ \ \ \ \sum\limits_{k = 1}^K {{\delta _{n,k}}}  = {K_n}, \ n \in {\cal N}, \label{original Constraints f}\\
 &  \ \ \ \ \ \sum\limits_{n = 1}^N {{\delta _{k,n}}}  = 1, \ k \in {\cal K},  \label{original Constraints g}\\
  & {\ \ \ \ \ \pi _n} \in  \Omega, \ n \in {\cal N},\label{original Constraints h}
\end{align}
\end{subequations}
where $\boldsymbol{\delta}  = \left\{ {{\delta _{1,1}}, \cdots ,{\delta _{N,K}}} \right\}$ is the channel assignment indication vector,  $\textbf{p} = \left\{ {{p_{1,1}}, \cdots ,{p_{N,K}}} \right\}$ is the power allocation vector, $\boldsymbol{\pi}  = \left\{ {{\pi _1}\left( 1 \right), \cdots ,{\pi _N}\left( K \right)} \right\}$ is the decoding order vector. Constraint \eqref{original Constraints b} guarantees the success of the SIC decoding. Constraint \eqref{original Constraints c} describes the minimum capacity requirement $R_{\min }$ of each user. Constraint \eqref{original Constraints d} indicates that the total transmit power budget is ${P_{\max }}$. Constraint \eqref{original Constraints e} is for the IRS reflection coefficients. Constraint \eqref{original Constraints f} demonstrates that each channel can be assigned to ${K_n}$ users. Constraint \eqref{original Constraints g} indicates that each user can be allocated to no more than one channel. In constraint \eqref{original Constraints h}, $\Omega$ is the combination set of all possible decoding orders.
\begin{theorem}
(P1) is a NP hard problem even when only the channel assignment problem is considered.

Proof: See Appendix A.
\end{theorem}

There are three main challenges to solve (P1). Firstly, due to the binary constraint of the indication vector, (P1) is a NP-hard problem. Secondly, since the decoding order can be controlled by the IRS reflection coefficients, it is difficult to obtain the optimal decoding order for the NOMA users. Thirdly, the transmit power and reflection coefficients are highly coupled, which makes the problem even more challenging. To find a feasible solution, we propose a three-step algorithm.
\section{The Proposed Algorithms for IRS-NOMA Systems}
To make (P1) tractable, we decouple the problem into three steps. Firstly, we fix the channel assignment and decoding order, then solve the joint power allocation and reflection coefficients design problem. For the channel assignment problem, a novel channel assignment algorithm based on many-to-one matching is proposed. In addition, to reduce the complexity of searching for the optimal decoding order, a low-complexity decoding order optimization algorithm is proposed.
\subsection{Joint Power Allocation and Reflection Coefficient Design}
Let ${k_n}$ denote the \emph{k}-th decoded user index on the \emph{n}-th channel. For a given channel assignment and decoding order, the capacity ${R_{n,{k_n}}}$ can be rewritten as
\begin{align}\label{Rate_n_k}
 {R_{n,{k_n}}} = \log \left( {1 + \frac{{{p_{n,{k_n}}}{{\left| {\textbf{g}_{n,{k_n}}^H\boldsymbol{\Theta}{\textbf{f}_n} + {h_{n,{k_n}}}} \right|}^2}}}{{{{\left| {\textbf{g}_{n,{k_n}}^H\boldsymbol{\Theta} {\textbf{f}_n} + {h_{n,{k_n}}}} \right|}^2}{P_{n,{k_n}}}  + {\sigma ^2}}}} \right),
\end{align}
where ${P_{n,{k_n}}} = \sum\limits_{{i_n} = {k_n} + 1}^{{K_n}} {{p_{n,{i_n}}}}$.

Furthermore, constraint \eqref{original Constraints b} in (P1) can be simplified as
 \begin{align}\label{rewritten decoding constraints}
 {\left| {\textbf{g}_{n,{{\overline k }_n}}^H\boldsymbol{\Theta} {\textbf{f}_n} + {h_{n,{{\overline k }_n}}}} \right|^2} - {\left| {\textbf{g}_{n,{k_n}}^H\boldsymbol{\Theta} {\textbf{f}_n} + {h_{n,{k_n}}}} \right|^2} \ge 0, \ {\overline k _n} > {k_n},
\end{align}
where ${k_n}, \ {\overline k _n} \in {{\cal K}_n}, \ n \in {\cal N}$.
To tackle the non-concavity of the objective function in (P1), we introduce the new variable set $\boldsymbol{\chi} {\rm{ = }}\left\{ {{\chi _{1,{k_1}}}, \cdots ,{\chi _{N,{K_N}}}} \right\}$,  whose elements satisfy the following inequality
 \begin{align}\label{objective inequaltiy}
 \frac{{{p_{n,{k_n}}}{{\left| {\textbf{g}_{n,{k_n}}^H\boldsymbol{\Theta} {\textbf{f}_n} + {h_{n,{k_n}}}} \right|}^2}}}{{{{\left| {\textbf{g}_{n,{k_n}}^H\boldsymbol{\Theta} {\textbf{f}_n} + {h_{n,{k_n}}}} \right|}^2}\sum\limits_{{i_n} = {k_n} + 1}^{{K_n}} {{p_{n,{i_n}}}}  + {\sigma ^2}}} \ge  {\chi _{n,{k_n}}}.
\end{align}

Then, (P1) can be equivalently transformed into the following problem
\begin{subequations}
\begin{align}\label{equivalent original problem}
  &{(\rm \textbf{P2})} \ \   \mathop {\max }\limits_{\textbf{p},\boldsymbol{\Theta} ,\boldsymbol{\chi} } \sum\limits_{n = 1}^N {\sum\limits_{k_n = 1}^{K_n} {{{\log }_2}\left( {1 + {\chi _{n,{k_n}}}} \right)} },  \\
  &{\rm s.t.} \  {\log _2}\left( {1 + {\chi _{n,{k_n}}}} \right) \ge R_{\min }, \ {k_n} \in {{\cal K}_n}, \ n \in {\cal N}, \label{equivalent original constraint b}\\
  & \ \ \ \ \ \sum\limits_{n = 1}^N {\sum\limits_{{k_n} = 1}^{{K_n}} {{p_{n,{k_n}}}} }  \le {P_{\max }}, \label{equivalent original constraint c}\\
  & \ \ \ \ \ \ \eqref{original Constraints e}, ~\eqref{rewritten decoding constraints}, ~\eqref{objective inequaltiy}. \label{equivalent original constraint d}
\end{align}
\end{subequations}

Since the variables $\textbf{p}$ and $\boldsymbol{\Theta}$ are coupled, (P2) is non-convex and difficult to be solved directly. To make (P2) tractable, we first divide it into the following two subproblems
\begin{subequations}
\begin{align}\label{power problem}
     & {(\rm \textbf{P2.1})} \ \   \mathop {\max }\limits_{\textbf{p},\boldsymbol{\chi} } \sum\limits_{n = 1}^N {\sum\limits_{k = 1}^K {{{\log }_2}\left( {1 + {\chi _{n,{k_n}}}} \right)} },  \\
      &{\rm s.t.} \ \eqref{objective inequaltiy}, ~\eqref{equivalent original constraint b}, ~\eqref{equivalent original constraint c}, \label{power problem c}
\end{align}
\end{subequations}
and
\begin{subequations}
\begin{align}\label{phase problem}
      & {(\rm \textbf{P2.2})} \ \    {\rm Find} \  \boldsymbol{\Theta},  \\
     &{\rm s.t.}\ ~\eqref{original Constraints e}, ~\eqref{rewritten decoding constraints}, ~\eqref{objective inequaltiy},\label{phase problem b}
\end{align}
\end{subequations}
 where subproblem (P2.1) focuses on finding the optimal power allocation vector \textbf{p} and subproblem (P2.2) focuses on finding the optimal reflection coefficient matrix $\boldsymbol{\Theta}$.

In the following, we discuss how to solve the above two subproblems.
\subsubsection{ Proposed Algorithm to Solve Subproblem (P2.1)} \ \

Before solving subproblem (P2.1), we rewrite the constraint \eqref{objective inequaltiy} as
 \begin{align}\label{rearrange objective inequaltiy}
{p_{n,{k_n}}} \ge {\chi _{n,{k_n}}}\sum\limits_{{i_n} = {k_n} + 1}^{{K_n}} {{p_{n,{i_n}}}}  + {\chi _{n,{k_n}}}{\nu _{n,{k_n}}},
\end{align}
where ${\nu _{n,{k_n}}}{\rm{ = }}{{{\sigma ^2}} \mathord{\left/
 {\vphantom {{{\sigma ^2}} {{{\left| {\textbf{g}_{n,{k_n}}^H\boldsymbol{\Theta} {\textbf{f}_n} + {h_{n,{k_n}}}} \right|}^2}}}} \right.
 \kern-\nulldelimiterspace} {{{\left| {\textbf{g}_{n,{k_n}}^H\boldsymbol{\Theta} {\textbf{f}_n} + {h_{n,{k_n}}}} \right|}^2}}}$.

 Since the first right hand term in inequality \eqref{rearrange objective inequaltiy} is quasi-concave, the constraint \eqref{rearrange objective inequaltiy} is still non-convex. Here, we use the convex upper bound approximation~\cite{tran2012fast} to deal with the non-convexity. Define $g\left( {x,y} \right) = xy$ and $f\left( {x,y} \right) = \frac{\alpha }{2}{x^2} + \frac{1}{{2\alpha }}{y^2}\left( {\alpha  > 0} \right)$, then $f\left( {x,y} \right)$ is always an upper bound on $g\left( {x,y} \right)$, i.e., $f\left( {x,y} \right) \ge g\left( {x,y} \right)$. Obviously, $f\left( {x,y} \right)$ is convex. When $\alpha  = \frac{y}{x}$, we have:$f\left( {x,y} \right) = g\left( {x,y} \right)$ and $\nabla f\left( {x,y} \right) = \nabla g\left( {x,y} \right)$, where $\nabla f\left( {x,y} \right)$ is the gradient of the function $f\left( {x,y} \right)$. Based on the above analysis, we have
\begin{align}\label{convex approximation}
 \begin{array}{l}
{\chi _{n,{k_n}}}\sum\limits_{{i_n} = {k_n} + 1}^{{K_n}} {{p_{n,{i_n}}}}  \le \frac{1}{2}\frac{1}{{{\alpha _{n,{k_n}}}}}{\left( {\sum\limits_{{i_n} = {k_n} + 1}^{{K_n}} {{p_{n,{i_n}}}} } \right)^2} + \frac{1}{2}{\alpha _{n,{k_n}}}\chi _{n,{k_n}}^2,
\end{array}
\end{align}
where ${{\alpha _{n,{k_n}}}}$ is a fixed point, ${\;{k_n} \in {\mathcal{K}_n},\;n \in \mathcal{N}}$.

Equality~\eqref{convex approximation} will always hold if ${\alpha _{n,{k_n}}}{\rm{ = }}\frac{{\sum\nolimits_{{i_n} = {k_n} + 1}^{{K_n}} {{p_{n,{i_n}}}} }}{{{\chi _{n,{k_n}}}}}$, ${{k_n} \in {\mathcal{K}_n},\;n \in \mathcal{N}}$. The fixed point ${\alpha _{n,{k_n}}}$ can be updated in the ${t_1}$-th iteration as follows
\begin{align}\label{update alpha }
 {\alpha _{n,{k_n}}}\left( {{t_1}} \right) = \frac{{\sum\nolimits_{{i_n} = {k_n} + 1}^{{K_n}} {{p_{n,{i_n}}}\left( {{t_1} - 1} \right)} }}{{{\chi _{n,{k_n}}}\left( {{t_1} - 1} \right)}}.
\end{align}

Then, constraint \eqref{objective inequaltiy} is approximated as
\begin{align}\label{convex appproximation}
{p_{n,{k_n}}} \ge \frac{1}{2}\frac{1}{{{\alpha _{n,{k_n}}}\left( {{t_1}{\rm{ - }}1} \right)}}{\left( {\sum\limits_{{i_n} = {k_n} + 1}^{{K_n}} {{p_{n,{i_n}}}} } \right)^2}{\rm{ + }}\frac{1}{2}{\alpha _{n,{k_n}}}\left( {{t_1}{\rm{ - }}1} \right)\chi _{n,{k_n}}^2 + {\chi _{n,{k_n}}}{\nu _{n,{k_n}}}.
\end{align}

Finally, solving subproblem (P2.1) is transformed to solving the following problem iteratively
\begin{subequations}
\begin{align}\label{OP20}
     &{(\rm \textbf{P3})} \ \     \mathop {\max }\limits_{\textbf{p},\boldsymbol{\chi} } \sum\limits_{n = 1}^N {\sum\limits_{k = 1}^K {{{\log }_2}\left( {1 + {\chi _{n,{k_n}}}} \right)} },    \\
      &{\rm s.t.} \ ~\eqref{equivalent original constraint b}, ~\eqref{equivalent original constraint c}, ~\eqref{convex appproximation}.
\end{align}
\end{subequations}

It is noted that (P3) is convex and can be solved efficiently by standard algorithms or software, such as CVX~\cite{grant2014cvx}. The proposed iterative power allocation algorithm to solve subproblem (P2.1) is summarized in \textbf{ Algorithm 1}.
\begin{algorithm}
\caption{Power Allocation Algorithm}
\label{Power Allocation Algorithm}
\begin{algorithmic}[1]
  \STATE  \textbf{Initialize} feasible points ${p_{n,{k_n}}}\left( 0 \right)$ and ${\chi _{n,{k_n}}}\left( 0 \right)$, ${k_n} \in {{\cal K}_n}, \ n \in {\cal N}$. Let iteration index ${t_1} = 1$.
\REPEAT
  \STATE    calculate ${\alpha _{n,{k_n}}}\left( {{t_1}} \right)$ according to~\eqref{update alpha }, ${k_n} \in {{\cal K}_n}, \ n \in {\cal N}$;
  \STATE   solve ({\rm P}3) to obtain ${p_{n,{k_n}}}\left( {{t_1}} \right)$ and ${\chi _{n,{k_n}}}\left( {{t_1}} \right)$, ${k_n} \in {{\cal K}_n}, \ n \in {\cal N}$;
  \STATE   ${t_1} = {t_1} + 1$;
\UNTIL { the objective value of (P2.1) converge}.
  \STATE  \textbf{Output}: optimal  $\textbf{p}$ and $\boldsymbol{\chi}$.
\end{algorithmic}
\end{algorithm}
\begin{remark}\label{remark:power algorithm}
Since the system throughput is upper bounded by a finite value and the objective value sequence of subproblem (P2.1) produced by \textbf{Algorithm 1} is non decreasing, the proposed iterative power allocation algorithm is guaranteed to converge.
\end{remark}

In \textbf{Algorithm 1}, the initial feasible points ${p_{n,{k_n}}}\left( 0 \right)$ and ${\chi _{n,{k_n}}}\left( 0 \right)$ are needed. Usually, it is difficult to find the feasible points. In the following, we formulate a feasibility problem and propose a novel feasible initial points searching algorithm. By introduce an infeasibility indicator $z\ge 0$, the feasibility problem in the ${t_2}$-th iteration is given as
\begin{subequations}
\begin{align}\label{feasible problem}
     &{(\rm \textbf{P4})} \ \      \mathop {\min }\limits_{\textbf{p},\boldsymbol{\chi},z } z,    \\
      &{\rm s.t.} \ {\log _2}\left( {1 + {\chi _{n,{k_n}}}} \right){\rm{ + }}z \ge R_{\min },\ {k_n} \in {{\cal K}_n}, \ n \in {\cal N}, \\
     & \ \ \ \ \begin{array}{l}
{p_{n,{k_n}}}{\rm{ + }}z \ge \frac{1}{2}\frac{1}{{{\alpha _{n,{k_n}}}\left( {{t_2}{\rm{ - }}1} \right)}}{\left( {\sum\limits_{{i_n} = {k_n} + 1}^{{K_n}} {{p_{n,{i_n}}}} } \right)^2}\\
\ \ \ \ \ \ \ \ \ \ \ {\rm{ + }}\frac{1}{2}{\alpha _{n,{k_n}}}\left( {{t_2}{\rm{ - }}1} \right)\chi _{n,{k_n}}^2 + {\chi _{n,{k_n}}}{\nu _{n,{k_n}}}, \ {k_n} \in {{\cal K}_n}, \ n \in {\cal N},
\end{array}  \\
    & \ \ \ \ \   \sum\limits_{n = 1}^N {\sum\limits_{{k_n} = 1}^{{K_n}} {{p_{n,{k_n}}}} }  \le {P_{\max }}{\rm{ + }}z,\ {k_n} \in {{\cal K}_n}, \ n \in {\cal N},
\end{align}
\end{subequations}
where $z$ denotes how far the corresponding constrains in (P3) are from being satisfied.

(P4) is also a convex optimization problem, which can be solved similarly as \textbf{Algorithm 1}. The proposed feasible points searching algorithm is summarized in\textbf{ Algorithm 2}.
\begin{algorithm}
\caption{Feasible Initial Points Searching Algorithm}
\label{Feasible Points Searching Algorithm}
\begin{algorithmic}[1]
  \STATE  Randomly initialize points ${p_{n,{k_n}}}\left( 0 \right)$ and ${\chi _{n,{k_n}}}\left( 0 \right)$, ${k_n} \in {{\cal K}_n}, \ n \in {\cal N}$. Let iteration index ${t_2} = 1$.
\REPEAT
  \STATE  calculate ${\alpha _{n,{k_n}}}\left( {{t_2}} \right)$ according to~\eqref{update alpha }, ${k_n} \in {{\cal K}_n},\ n \in {\cal N}$;
  \STATE   solve (P4) to obtain ${p_{n,{k_n}}}\left( {{t_2}} \right)$ and ${\chi _{n,{k_n}}}\left( {{t_2}} \right)$, ${k_n} \in {{\cal K}_n},\ n \in {\cal N}$;
  \STATE   ${t_2} = {t_2} + 1$;
\UNTIL {\emph{z} below a threshold $\xi  > 0$}.
  \STATE   \textbf{Output}: optimal $\textbf{p}$ and $\boldsymbol{\chi}$.
\end{algorithmic}
\end{algorithm}
\begin{remark}\label{remark:algorithm 4}
 Different from \textbf{Algorithm 1}, ${p_{n,{k_n}}}\left( 0 \right)$ and ${\chi _{n,{k_n}}}\left( 0 \right)$ in \textbf{Algorithm 2} can be initialized randomly. When $z = 0$, the optimal solutions of (P4) are feasible for (P3). Therefore, the output of \textbf{Algorithm 2} can be used to replace the initial feasible points ${p_{n,{k_n}}}\left( 0 \right)$ and ${\chi _{n,{k_n}}}\left( 0 \right)$ in \textbf{Algorithm 1}.
\end{remark}
\subsubsection{ Proposed Algorithm to Solve Subproblem (P2.2)} \ \

The combined channel gain ${\left| {\textbf{g}_{n,k_n}^H\boldsymbol{\Theta} {\textbf{f}_n} + {h_{n,k_n}}} \right|^2}$ can be reformulated as
\begin{align}\label{reformulated combined channel gains}
 {\left| {\textbf{g}_{n,k_n}^H\boldsymbol{\Theta} {\textbf{f}_n} + {h_{n,k_n}}} \right|^2} = {\left| {{\textbf{z}_{n,{k_n}}}\boldsymbol{e_\theta } + {h_{n,{k_n}}}} \right|^2},
\end{align}
where ${\textbf{z}_{n,{k_n}}} = \textbf{g}_{n,{k_n}}^H {\rm diag}\left\{ {{\textbf{f}_n}} \right\}$ and ${\textbf{e}_\theta } = \left[ {{\lambda _1}{e^{j{\theta _1}}}_{}{\lambda _2}{e^{j{\theta _2}}}_{}{ \cdots _{}}{\lambda _M}{e^{j{\theta _M}}}} \right]^T$.

We introduce variables ${\kappa _{n,{k_n}}}$ and ${\xi _{n,{k_n}}}$, which are defined as
\begin{align}\label{real part }
{\kappa _{n,{k_n}}}{\rm{ = }} {\rm {real}}\left( {{\textbf{z}_{n,{k_n}}}\boldsymbol{e_\theta } + {h_{n,{k_n}}}} \right),
\end{align}
\vspace{-1cm}
\begin{align}\label{imag part}
{\xi _{n,{k_n}}}{\rm{ = }}{\rm {imag}}\left( {{\textbf{z}_{n,{k_n}}}{\boldsymbol{e_\theta} } + {h_{n,{k_n}}}} \right),
\end{align}
where $\kappa _{n,{k_n}}^2 + \xi _{n,{k_n}}^2={\left| {{\textbf{z}_{n,{k_n}}}\boldsymbol{e_\theta } + {h_{n,{k_n}}}} \right|^2} $, ${k_n} \in {{\cal K}_n},\ n \in {\cal N}$.

Then, subproblem (P2.2) can be rewritten as
\begin{subequations}
\begin{align}\label{theta problem}
     &(\rm \textbf{P5}) \ \      {\rm Find} \quad  \boldsymbol{e_\theta },  \\
      &{\rm s.t.} \  \kappa _{n,{{\overline k }_n}}^{^2}{\rm{ + }}\zeta _{n,{{\overline k }_n}}^{^2} > \kappa _{n,{k_n}}^{^2}{\rm{ + }}\zeta _{n,{k_n}}^{^2},\ {\overline k _n} > {k_n},\label{theta problem b} \\
     & \ \ \ \ \ \kappa _{n,{k_n}}^{^2}{\rm{ + }}\zeta _{n,{k_n}}^{^2} \ge \left( {\kappa _{n,{k_n}}^{^2}{\rm{ + }}\zeta _{n,{k_n}}^{^2}} \right){\beta _{n,{k_n}}}{\rm{ + }}{\chi _{n,{k_n}}}{\sigma ^2},\label{theta problem c}  \\
     & \ \  \ \ \left| {\boldsymbol{e_\theta }\left( m \right)} \right| \le 1, \ m \in {\cal M},\label{theta problem d}  \\
      & \  \ \ \  ~\eqref{real part },~\eqref{imag part},\label{theta problem e}
\end{align}
\end{subequations}
where ${\beta _{n,{k_n}}}{\rm{ = }}\frac{{{\chi _{n,{k_n}}}\sum\nolimits_{{i_n} = {k_n} + 1}^{{K_n}} {{p_{n,{i_n}}}} }}{{{p_{n,{k_n}}}}}$, ${k_n},\ {\overline k _n} \in {{\cal K}_n},\ n \in {\cal N}$.

(\rm P5) is still a non-convex problem, due to the non-convex constraints~\eqref{theta problem b} and~\eqref{theta problem c}. To deal with the non-convexity, the SCA method can be used. At point $\left( {{{\widetilde \kappa }_{n,{k_n}}},{{\widetilde \xi }_{n,{k_n}}}} \right)$, the first-order approximation of $\kappa _{n,{k_n}}^{^2}{\rm{ + }}\zeta _{n,{k_n}}^{^2}$  is
\begin{align}\label{first-order approximation}
 \kappa _{n,{k_n}}^{^2}{\rm{ + }}\zeta _{n,{k_n}}^{^2} \ge \widetilde \kappa _{n,{k_n}}^2{\rm{ + }}\widetilde \zeta _{n,{k_n}}^2{\rm{ + }}2{\widetilde \kappa _{n,{k_n}}}\left( {{\kappa _{n,{k_n}}} - {{\widetilde \kappa }_{n,{k_n}}}} \right)  +2{\widetilde \zeta _{n,{k_n}}}\left( {{\zeta _{n,{k_n}}} - {{\widetilde \zeta }_{n,{k_n}}}} \right) {\rm{ = }}{\varphi _{n,{k_n}}}\left( {{\kappa _{n,{k_n}}},{\zeta _{n,{k_n}}}} \right),
 \end{align}
where the point $\left( {{{\widetilde \kappa }_{n,{k_n}}},{{\widetilde \xi }_{n,{k_n}}}} \right)$ can be updated in the ${t_3}$-th iteration as
\begin{align}\label{update kappa}
{\widetilde \kappa _{n,{k_n}}}\left( {{t_3}} \right){\rm{ = }}{\rm real}\left( {{\textbf{z}_{n,{k_n}}}\boldsymbol{e_\theta }\left( {{t_3} - 1} \right) + {h_{n,{k_n}}}} \right),
\end{align}
\vspace{-1cm}
\begin{align}\label{update xi}
{\widetilde \xi _{n,{k_n}}}\left( {{t_3}} \right){\rm{ = }}{\rm imag}\left( {{\textbf{z}_{n,{k_n}}}\boldsymbol{e_\theta }\left( {{t_3} - 1} \right) + {h_{n,{k_n}}}} \right).
\end{align}

Thus, constraints ~\eqref{theta problem b} and ~\eqref{theta problem c} can be approximated, respectively, as
\begin{align}\label{approximated 1}
 {\varphi _{n,{{\overline k }_n}}}\left( {{\kappa _{n,{{\overline k }_n}}},{\zeta _{n,{{\overline k }_n}}}} \right) > \kappa _{n,{k_n}}^{^2}{\rm{ + }}\zeta _{n,{k_n}}^{^2},
\end{align}
\vspace{-1cm}
\begin{align}\label{approximated 2}
 {\varphi _{n,{k_n}}}\left( {{\kappa _{n,{k_n}}},{\zeta _{n,{k_n}}}} \right) \ge \left( {\kappa _{n,{k_n}}^{^2}{\rm{ + }}\zeta _{n,{k_n}}^{^2}} \right){\beta _{n,{k_n}}}{\rm{ + }}{\chi _{n,{k_n}}}{\sigma ^2},
\end{align}
where ${\overline k _n} > {k_n}$ and ${k_n},\ {\overline k _n} \in {{\cal K}_n},\ n \in {\cal N}$.

Consequently, solving subproblem (P2.2) is transformed to iteratively solving the following problem
\begin{subequations}
\begin{align}\label{zeta problem}
     & {(\rm \textbf{P6})} \ \    {\rm Find} \quad  \boldsymbol{e_\theta },   \\
      &{\rm s.t.}   ~\eqref{real part }, ~\eqref{imag part}, ~\eqref{theta problem d}, ~\eqref{approximated 1}, ~\eqref{approximated 2}.
\end{align}
\end{subequations}

(\rm P6) is a convex optimization problem, which can be solved efficiently using CVX~\cite{grant2014cvx}. The proposed iterative reflection coefficients design algorithm to solve subproblem (P2.2) is summarized in \textbf{Algorithm 3}.
\begin{algorithm}
\caption{Reflection Coefficients Design Algorithm}
\label{reflection coefficient Optimization Algorithm}
\begin{algorithmic}[1]
  \STATE  \textbf{Initialize} $\boldsymbol{e_\theta }\left( 0 \right)$ and let iteration index ${t_3} = 1$.
  \REPEAT
  \STATE  update ${\widetilde \kappa _{n,{k_n}}}\left( {{t_3}} \right)$ and ${\widetilde \xi _{n,{k_n}}}\left( {{t_3}} \right)$ according to~\eqref{update kappa} and~\eqref{update xi}, respectively, ${k_n} \in {{\cal K}_n},\ n \in {\cal N}$;
  \STATE   solve (\rm P6) to obtain $\boldsymbol{e_\theta }\left( t_3 \right)$, ${\kappa _{n,{k_n}}}\left( {{t_3}} \right)$ and ${\xi _{n,{k_n}}}\left( {{t_3}} \right)$,  ${k_n} \in {{\cal K}_n},\ n \in {\cal N}$;
  \STATE   ${t_3} = {t_3} + 1$;
  \UNTIL { $\boldsymbol{e_\theta }\left( t_3 \right)$, ${\kappa _{n,{k_n}}}\left( {{t_3}} \right)$ and ${\xi _{n,{k_n}}}\left( {{t_3}} \right)$ converge}.
  \STATE   \textbf{Output}: optimal $\boldsymbol{e_\theta }$.
\end{algorithmic}
\end{algorithm}
\subsection{Channel Assignment Algorithm based on Many-to-One Matching}
In this subsection, we solve the channel assignment problem. Assume that the power allocation and reflection coefficients are fixed, then (P1) can be reformulated as
 \begin{subequations}
\begin{align}\label{OP6}
  & {(\rm \textbf{P7})} \ \   \mathop {\max }\limits_{\boldsymbol{\delta}}  \sum\limits_{n = 1}^N {\sum\limits_{k = 1}^K {{R_{n,k}}} },  \\
  &{\rm s.t.} \  ~\eqref{original Constraints b}, ~\eqref{original Constraints c}, ~\eqref{original Constraints f}, ~\eqref{original Constraints g}.
\end{align}
\end{subequations}

The above problem can be solved by many-to-one matching with two sides, i.e., channels and users. Combining with the channel assignment problem, define the many-to-one matching function $\Upsilon$ as~\cite{cui2017optimal,zhao2017spectrum}
\begin{enumerate}
  \item $\left| {\Upsilon \left( k \right)} \right| = 1$, $\forall k \in {\cal K}$, $\Upsilon \left( k \right) \in {\cal N}$;
  \item $\left| {\Upsilon \left( n \right)} \right| = {K_n}$, $n \in {\cal N}$;
  \item $\Upsilon \left( k \right){\rm{ = }}n$ if and only if $k \in \Upsilon \left( n \right)$,
\end{enumerate}
where definition 1) means that each user can only be matched with one channel; definition 2) provides the maximum number of users that can be allocated to each channel; definition 3) implies that if user \emph{k} is matched with channel \emph{n}, then channel \emph{n} is also matched with user \emph{k}.

 Define the utility functions of user \emph{k} and channel \emph{n} as: ${U_{n,k}} = {R_{n,k}}$ and ${U_n} = \sum\nolimits_{k \in \Upsilon \left( n \right)} {{U_{n,k}}}$, respectively. Since the utility of user \emph{k} depends not only on the channel it is allocated to but also on the set of users in the same channel. To tackle this interdependence, we utilize swap operations between any two users to exchange their allocated channels. First, define swap matching~\cite{cui2017optimal,zhao2017spectrum} as follows
\begin{align}\label{define swap matching}
 \Upsilon _k^{\widetilde k} = \left\{ {\Upsilon \backslash \left\{ {\left( {k,n} \right),\left( {\widetilde k,\widetilde n} \right)} \right\} \cup \left\{ {\left( {\widetilde k,n} \right),\left( {k,\widetilde n} \right)} \right\}} \right\},
\end{align}
where $\Upsilon \left( k \right) = n$ and $\Upsilon \left( {\widetilde k} \right) = \widetilde n$.

The swap matching enables user \emph{k} and user ${\widetilde k}$ to switch their assigned channels. Then, we introduce the definition of swap-blocking pair. Given a matching function $\Upsilon$ and assume that $\Upsilon \left( k \right) = n$ and $\Upsilon \left( {\widetilde k} \right) = \widetilde n$, a pair of users $\left( {k,\widetilde k} \right)$ is a swap-blocking pair if and only if
\begin{enumerate}
  \item $\forall \omega  \in \left\{ {k,\widetilde k,n,\widetilde n} \right\}$, ${U_\omega }\left( {\Upsilon _k^{\widetilde k}} \right) \ge {U_\omega }\left( \Upsilon  \right)$;
  \item $\exists \omega  \in \left\{ {k,\widetilde k,n,\widetilde n} \right\}$, ${U_\omega }\left( {\Upsilon _k^{\widetilde k}} \right) > {U_\omega }\left( \Upsilon  \right)$,
\end{enumerate}
where ${U_\omega }\left( \Upsilon  \right)$ is the utility of player $\omega$ (user $\omega$ or channel $\omega$), under the matching state $\Upsilon$.

According to the above definition, it is noted that if two users want to switch their assigned channels, both of the conditions should be satisfied. Condition 1) indicates that all the involved players' utilities should not be reduced after the swap operation; Condition 2) indicates that after the swap operation, at least one of the players' utilities is increased.

Based on the above analysis, the proposed channel assignment algorithm is summarized in \textbf{Algorithm 4}.  There are two processes in \textbf{Algorithm 4} as follows
\begin{enumerate}
  \item \textbf{Initialization Process}: The set of users assigned to channel \emph{n} is denoted as ${{\cal K}_n}$, the set of users rejected by the \emph{n}-th channel is denoted as ${{\cal R}_n}$ and the set of channels rejected by user \emph{k} is ${\widetilde {\cal R}_k}$, the set of users that are not matched with any channel is denoted as ${{\cal K}_{\rm NOT}}$. Denote the set of users that propose to the \emph{n}-th channel as ${\cal K}_n^{\rm PRO}$ and define the user set ${{\cal T}_n} = {{\cal K}_n} \cup {\cal K}_n^{\rm PRO}$. Let \emph{Q} be the total number of users in set ${{\cal T}_n}$ and $k_q^n \ \left( {q = 1, \cdots ,Q} \right)$ be the \emph{q}-th user in set ${{\cal T}_n},\ n \in {\cal N}$. During the matching period, each un-matched user proposes to the channel that can provide the highest equivalent channel gain and has never rejected it before. Then each channel accepts the proposal with the highest channel gain it can provide and rejects other users. Repeat the above process until the set of un-matched users is empty.
  \item \textbf{Swapping Process}: Swap operations among users are enabled to further improve the performance of the channel assignment algorithm. With the obtained user set ${\cal K}_n$, $n \in {\cal N}$, in the \textbf{Initialization Process}, each user tries to search for another user to construct the swap-blocking pair and update their corresponding matching state and user set ${\cal K}_n$, $n \in {\cal N}$. This operation will continue until there is no swap-blocking pair.
\end{enumerate}
\begin{theorem}
The proposed channel assignment algorithm in \textbf{Algorithm 4} converges to a two-sided stable matching within a limited number of iterations.

Proof: See Appendix B.
\end{theorem}
\begin{algorithm}[!t]
\caption{ Channel Assignment Algorithm}
\label{Algorithm 5}
\begin{algorithmic}[1]
  \STATE \textbf{Initialization Process}:
  \STATE   Initialize ${{\cal K}_n} = \emptyset ,{{\cal R}_n} = \emptyset ,{\widetilde {\cal R}_k} = \emptyset$ and ${\cal
               K}_n^{\rm PRO} = \emptyset \left( {n \in {\cal N},k \in {\cal K}} \right)$ and set ${{\cal K}_{\rm NOT}} = {\cal K}$.
  \WHILE {${{\cal K}_{\rm NOT}} \ne \emptyset$}
  \STATE the un-matched user $k \in \left\{ {{{\cal K}_{\rm NOT}}\setminus{ \cup _{n \in {\cal N}}}{{\cal N}_n}} \right\}$ proposes to choose its best channel \emph{n}, where $n = \arg \mathop {\max }\limits_{n \in \left\{ {{\cal N}\backslash
               {{\widetilde {\cal R}}_k}} \right\}} {\left| {\textbf{g}_{n,k}^H\boldsymbol{\Theta} {\textbf{f}_n} + {h_{n,k}}} \right|^2}$;
  \STATE  update ${\cal K}_n^{\rm PRO}$ based on the results obtained from the last step, $ n \in {\cal  N}$;
  \STATE  update set ${{\cal T}_n} = \left\{ {k_1^n, \cdots ,k_Q^n} \right\} = {{\cal K}_n} \cup {\cal K}_n^{\rm PRO}$, $ n \in {\cal  N}$;
  \IF {$Q \le {K_n}$, $n \in {\cal N}$}
  \STATE  the \emph{n}-th channel accepts all the users in ${{\cal T}_n}$, ${{\cal K}_n} = {{\cal T}_n}$;
   \ELSE
  \STATE  update ${{\cal K}_n} = \left\{ {k_1^n, \cdots ,k_{{K_n}}^n} \right\}$, where ${k_1^n, \cdots ,k_{{K_n}}^n}$ are the first ${K_n}$ largest ${\left| {\textbf{g}_{n,k}^H\boldsymbol{\Theta} {\textbf{f}_n} + {h_{n,k}}} \right|^2}$ in ${{\cal T}_n}$;
  \STATE  update ${{\cal R}_n} = {{\cal R}_n} \cup \left\{ {k_{{K_n} + 1}^n, \cdots ,k_Q^n} \right\}$
    \STATE  update ${\widetilde {\cal R}_k} = {\widetilde {\cal R}_k} \cup \left\{ {n|k \in {{\cal R}_n},n \in {\cal N}} \right\}$;
  \ENDIF
  \ENDWHILE
  \STATE \textbf{Swapping Process}:
  \STATE   For any user $k \in {{\cal K}_n}$, it searches for another user $\widetilde k \in {{\cal K}_{\widetilde n}}$, where
                 $\widetilde n \ne n,\ n \in {\cal N}$.
  \IF {user pair $\left( {k, \widetilde k} \right)$ is a swap-blocking pair}
  \STATE   update $n = \Upsilon \left( {\widetilde k} \right)$ and $\widetilde n = \Upsilon \left( k \right)$;
  \STATE   update ${{\cal K}_n}$ and ${{\cal K}_{\widetilde n}},\ \widetilde n \ne n,\ n \in {\cal N}$;
   \ELSE
  \STATE   keep the current matching state unchanged;
  \ENDIF
  \STATE  \textbf{Repeat} step 16 - step 22 until there is no swap-blocking pair.
  \STATE  \textbf{Output}: user set ${{\cal K}_n},\ n \in {\cal N}$.
\end{algorithmic}
\end{algorithm}
\subsection{A Low Complexity Scheme for Decoding Order Optimization}
In NOMA systems, the decoding order is important for canceling interference from the other users sharing the same channel~\cite{liu2017enhancing}. For the considered IRS-NOMA system, the SIC decoding order depends on
the combined channel gain of both the direct link and the reflection links, which are controlled by the IRS. The optimal decoding order in each channel will be any one of the ${{K_n}!}$ different decoding orders and (P1) must be solved ${{K_n}!}$ times. Therefore, an exhaustive search is needed over all the decoding orders which is highly complex. Here, we propose a low complexity decoding order optimization method by maximizing the sum of all the combined channel gains, which only needs to solve one optimization problem.
The formulated problem is as follows
\begin{subequations}
\begin{align}\label{solving decoding order}
     &{(\rm \textbf{P8})} \ \      \mathop {\max }\limits_{\boldsymbol{\Theta}}  \sum\limits_{n = 1}^N {\sum\limits_{{k} = 1}^{{K_n}} {{{\left| {\textbf{g}_{n,{k}}^H\boldsymbol{\Theta} {\textbf{f}_n} + {h_{n,{k}}}} \right|}^2}} },  \\
      &{\rm s.t.} \  \left| {\boldsymbol{\Theta}\left( m \right)} \right| \le 1,\ m \in {\cal M}.
\end{align}
\end{subequations}

Before solving (P8), we rewrite the combined channel gains ${\left| {\textbf{g}_{n,{k}}^H\boldsymbol{\Theta} {\textbf{f}_n} + {h_{n,{k_n}}}} \right|^2}$ as
\begin{align}
{\left| {\textbf{g}_{n,{k}}^H\boldsymbol{\Theta} {\textbf{f}_n} + {h_{n,{k}}}} \right|^2} = {\left| {{\textbf{v}_{n,{k}}}\textbf{e}} \right|^2}= {\rm Tr}\left( {{\textbf{V}_{{k},n}}\textbf{E}} \right),
\end{align}
where ${\textbf{v}_{n,{k}}} = \left[ {\textbf{g}_{n,{k}}^H {\rm diag}{{\left\{ {{\textbf{f}_n}} \right\}}_{}} \ \ {h_{n,{k}}}} \right]$, $\textbf{e} = {\left[ {{\lambda _1}{e^{j{\theta _1}}} \ {\lambda _2}{e^{j{\theta _2}}} \ \cdots \ {\lambda _M}{e^{j{\theta _M}}}_{} \ 1} \right]^T}$, ${\textbf{V}_{{k},n}} = \textbf{v}_{{k},n}^H{\textbf{v}_{{k},n}}$, $\textbf{E} = \textbf{e}{\textbf{e}^H}$ and the rank of matrix $\textbf{E}$ is one, i.e., ${\rm rank}\left( \textbf{E} \right) = 1$.

By exploiting the SDR, (P8) can be transformed into the following problem
\begin{subequations}
\begin{align}\label{relax problem}
     & {(\rm \textbf{P9})} \ \     \mathop {\max }\limits_{\textbf{E}} {\rm{Tr}}\left( {\sum\limits_{n = 1}^N {\sum\limits_{k = 1}^{{K_n}} {{{\textbf{V}}_{n,k}}} } {\textbf{E}}} \right), \\
      &{\rm s.t.} \  \textbf{E}\left( {m,m} \right) \le 1,\ m \in {\cal M},  \\
      &   \  \   \ \  \   \textbf{E}\left( {M{\rm{ + }}1,M{\rm{ + }}1} \right) = 1,  \\
       &   \  \  \   \   \  \textbf{E}\succeq \textbf{0}.
\end{align}
\end{subequations}

It is noted that (P9) is a standard semidefinite programming (SDP) problem, which can be solved by CVX~\cite{grant2014cvx}.

 (P9) is equivalent to (P8) if and only if the optimal solution ${\textbf{E}^{\rm{*}}}$ is a rank-one positive semidefinite matrix. However, the rank of the solution to (P9) may not be one because the rank-one constraint is relaxed. To solve this problem, a randomization method can be applied to construct a rank-one solution from the higher-rank ${\textbf{E}^{\rm{*}}}$. According to the Gaussian randomization method~\cite{yang2019intelligentNOMA}, we first calculate the eigen-decompostion of ${\textbf{E}^{\rm{*}}} = \textbf{U}{\boldsymbol{\Lambda}}{\textbf{U}^H}$, then generate \emph{L} candidate vectors as follows
\begin{align}\label{L candidate vectors}
 {\boldsymbol{\widetilde e}_l} = \textbf{U}{\boldsymbol{\Lambda} ^{\frac{1}{2}}}{\textbf{r}_l},\ l = 1,2 \cdots ,L,
\end{align}
where $\textbf{U}$ is the unitary matrix of eigenvectors, $\boldsymbol{\Lambda}$ denotes a diagonal matrix of eigenvalues, and ${\textbf{r}_l}$ is a random vector whose elements are independent random variables uniformly distributed on the unit circle in the complex plane.

Then, the following equality holds for any realization of ${\boldsymbol{\widetilde e}_l}$
 \begin{align}\label{realization of candidate vector}
{\left( {{{\boldsymbol{\widetilde e}}_l}} \right)^H}{\boldsymbol{\widetilde e}_l} = \textbf{r}_l^H{\left( {{\boldsymbol{\Lambda}  ^{\frac{1}{2}}}} \right)^H}{\textbf{U}^H}\textbf{U}{\boldsymbol{\Lambda}  ^{\frac{1}{2}}}{\textbf{r}_l} = {\rm Tr}\left( {\boldsymbol{\Lambda}  {\textbf{r}_l}\textbf{r}_l^H} \right)  = {\rm Tr}\left( \boldsymbol{\Lambda}  \right) ={\rm Tr}\left( {{\textbf{E}^{\rm{*}}}} \right).
\end{align}

With ${\boldsymbol{\widetilde e}_l}$, we can obtain the candidate reflection coefficient matrix as
 \begin{align}\label{realization of candidate vector}
 {\boldsymbol{\Theta} _l} = {\rm{diag}}\left\{ {{e^{j\angle \frac{{{{\boldsymbol{\widetilde e}}_l}\left[ 1 \right]}}{{{{\boldsymbol{\widetilde e}}_l}\left[ {M + 1} \right]}}}},{e^{j\angle \frac{{{{\boldsymbol{\widetilde e}}_l}\left[ 2 \right]}}{{{{\boldsymbol{\widetilde e}}_l}\left[ {M + 1} \right]}}}}, \cdots ,{e^{j\angle \frac{{{{\boldsymbol{\widetilde e}}_l}\left[ M \right]}}{{{{\boldsymbol{\widetilde e}}_l}\left[ {M + 1} \right]}}}}} \right\}.
\end{align}

The optimal reflection coefficient matrix selected from $\left\{ {{{\boldsymbol{\Theta} _l}}} \right\}$ satisfies all the constraints and maximizes the objective function value of (P8). The proposed decoding order optimization algorithm is summarized in \textbf{Algorithm~\ref{Decoding Order Optimization Algorithm}}.
\begin{algorithm}
\caption{Decoding Order Optimization Algorithm}
\label{Decoding Order Optimization Algorithm}
\begin{algorithmic}[1]
  \STATE  Solve (P9) to obtain ${\textbf{E}^{\rm{*}}}$;
\IF {${\rm rank}\left( {{\textbf{E}^{\rm{*}}}} \right) = 1$}
  \STATE    calculate nonzero eigenvalue ${\lambda _{\rm {eigen}}}$ of matrix ${\textbf{E}^{\rm{*}}}$  and its corresponding eigenvector ${\textbf{v}_{{\rm eigen}}}$ via eigen-decomposition;
   \STATE    calculate ${\boldsymbol{\Theta} ^*} = {\rm{ diag}}\left\{ {{\textbf{v}_{\rm {eigen}}}*\sqrt {{\lambda _{ \rm {eigen}}}} } \right\}$;
\ELSE
  \FOR { $l = 1,2, \cdots ,L$}
  \STATE    calculate ${\boldsymbol{\Theta} _l}$ according to ~\eqref{realization of candidate vector};
  \STATE    calculate the objective value of (P8);
  \ENDFOR
 \ENDIF
  \STATE  Let ${\boldsymbol{\Theta} ^*}{\rm{ = }}{\boldsymbol{\Theta} _{{l^*}}}$, where ${l^*} = \arg \mathop {\max }\limits_{l = 1,2, \cdots ,L} \sum\limits_{n = 1}^N {\sum\limits_{{k} = 1}^{{K_n}} {{{\left| {\textbf{g}_{n,{k}}^H{\boldsymbol{\Theta} _l}{\textbf{f}_n} + {h_{n,{k}}}} \right|}^2}} }$;
  \STATE  Calculate all combined channel gains $\left\{ {{\left| {\textbf{g}_{n,{k}}^H\boldsymbol{\Theta} {\textbf{f}_n} + {h_{n,{k}}}} \right|^2},\ k \in {{\cal K}_n},\ n \in {\cal N} } \right\}$ and rank them in ascending order for each channel;
  \STATE \textbf{Output}: decoding order ${\pi _n}\left( k \right),\ k \in {{\cal K}_n},\ n \in {\cal N}$.
\end{algorithmic}
\end{algorithm}
\subsection{Proposed Three-Step Resource Allocation Algorithm for IRS-NOMA Systems}
Based on the proposed \textbf{Algorithm 1} to \textbf{Algorithm 5} in the previous subsections, the proposed three-step optimization algorithm for the IRS-NOMA system is summarized in \textbf{Algorithm 6}. In the first step, channel assignment is performed based on \textbf{Algorithm 4}. In the second step, the SIC decoding orders for NOMA users in each channel are obtained according to the proposed low-complexity decoding order optimization algorithm, {\rm i.e.}, \textbf{Algorithm 5}. In the third step, the joint power allocation and reflection coefficients design algorithm is executed based on the channel assignment and decoding order optimization results obtained from the last two steps. Specifically, the feasible initial points searching algorithm, {\rm i.e.}, \textbf{Algorithm 2}, is first performed to get the initial points. Then, the power allocation algorithm, i.e., \textbf{Algorithm 1} and the reflection coefficients design algorithm, {\rm i.e.}, \textbf{Algorithm 3}, are performed alternatively until converge.
\begin{algorithm}
\caption{Proposed Three-Step Resource Allocation Algorithm for IRS-NOMA Systems}
\label{JCDPR-IRSNOMA}
\begin{algorithmic}[1]
\STATE -\textbf{Step 1}: Channel assignment
\STATE Obtain user index sets $\left\{ {{{\cal K}_n},\ n \in {\cal N}} \right\}$ via \textbf{Algorithm 4}.
\STATE -\textbf{Step 2}: SIC decoding order optimization
\STATE Obtain decoding orders $\left\{ {{\pi _n}\left( k \right),k \in {{\cal K}_n},n \in {\cal N}} \right\}$ via \textbf{Algorithm 5}.
\STATE -\textbf{Step 3}: Joint power allocation and reflection coefficient design
\STATE Randomly initialize $\textbf{e}_{\boldsymbol{\theta}} ^{\left( 0 \right)}$ and let iteration number ${t_0} = 1$.
\STATE Find feasible initial points ${\textbf{p}^{\left( 0 \right)}}$ and ${\boldsymbol{\chi} ^{\left( 0 \right)}}$ via \textbf{Algorithm 2}.
\REPEAT
    \STATE update ${\textbf{p}^{\left( {{t_0}} \right)}}$ and ${{\boldsymbol{\chi}}^{\left( {{t_0}} \right)}}$ via \textbf{Algorithm 1} with $\textbf{e}_{\boldsymbol{\theta}}^{\left( {{t_0}{\rm{ - }}1} \right)}$;
    \STATE update $\textbf{e}_{\boldsymbol{\theta}}^{\left( {{t_0}} \right)}$ via \textbf{Algorithm 3} with  ${\textbf{p}^{\left( {{t_0}} \right)}}$ and ${{\boldsymbol{\chi}}^{\left( {{t_0}} \right)}}$;
    \STATE ${t_0} = {t_0} + 1$
\UNTIL {the objective value of (P1) converges.}
\end{algorithmic}
\end{algorithm}
\subsection{Complexity and Convergence of the Proposed Three-Step Resource Allocation Algorithm}
\subsubsection{Complexity analysis} \ \

   The complexities of \textbf{Algorithm 1}, \textbf{Algorithm 2} and \textbf{Algorithm 3} with the interior-point method are $o\left( {8{K^3} + 2K\left( {2K{\rm{ + }}1} \right)} \right)$, $o\left( {{{\left( {2K + 1} \right)}^3} + {{\left( {2K + 1} \right)}^2}} \right)$ and $o\left\{ {{M^3} \\ + \left( {3K + \sum\limits_{n = 1}^N {\frac{{{K_n}\left( {{K_n}{\rm{ - }}1} \right)}}{2}} } \right)M} \right\}$, respectively. In \textbf{Algorithm 4}, the complexity of the initial process mainly depends on the number of users making proposals. In the worst case, the number of users making proposals is $K{N^2}$. In the second process, the maximum number of swap operations is ${K^2}$~\cite{zhao2017spectrum}. In \textbf{Algorithm 5}, the complexity of solving SDP problem is on the order of ${\left( {M{\rm{ + }}1} \right)^6}$.
  \subsubsection{Convergence analysis}\ \

 The convergence of the proposed three-step resource allocation algorithm mainly depends on Step 3. In the following, we will prove the convergence of the iterative procedure in Step 3. Let ${{\textbf{p}}^{{t_0}}},{{\boldsymbol{\chi}}^{{t_0}}}$ and $\textbf{e}_{\boldsymbol{\theta}}^{{t_0}}$ be the ${t_0}$-th iteration solution obtained by \textbf{Algorithm 6}. According to~\eqref{reformulated combined channel gains}, we have ${{\boldsymbol{\Theta }}^{{t_0}}} = {\rm diag}\left( {\textbf{e}_{\boldsymbol{\theta}} ^{{t_0}}} \right)$. Define ${R_{\rm P1}}\left( {{{\textbf{p}}^{{t_0}}},{{\boldsymbol{\Theta }}^{{t_0}}}} \right)$ and ${R_{\rm P2}}\left( {{{\textbf{p}}^{{t_0}}}, {{\boldsymbol{\chi}}^{{t_0}}}, \textbf{e}_{\boldsymbol{\theta}} ^{{t_0}}} \right)$ as the objective values of (P1) and (P2) in the ${t_0}$-th iteration, respectively. For (P2) with a given reflection coefficients vector $\textbf{e}_{\boldsymbol{\theta}} ^{{t_0} - 1}$,  we have the following inequality
 \begin{align}\label{budengshi1}
 {R_{\rm P1}}\left( {{{\textbf{p}}^{{t_0} - 1}},{{\boldsymbol{\Theta }}^{{t_0} - 1}}} \right)\mathop  = \limits^{\left( a \right)} {R_{\rm P2}}\left( {{{\textbf{p}}^{{t_0} - 1}},{{\boldsymbol{\chi}}^{{t_0} - 1}},\textbf{e}_{\boldsymbol{\theta}} ^{{t_0} - 1}} \right)\mathop  \le \limits^{\left( b \right)} {R_{\rm P2}}\left( {{{\textbf{p}}^{{t_0}}},{{\boldsymbol{\chi}}^{{t_0}}},\textbf{e}_{\boldsymbol{\theta}} ^{{t_0} - 1}} \right),
 \end{align}
where (a) comes from the fact that (P1) is equivalent to (P2) with optimal ${\boldsymbol{\chi}}$; (b) holds since ${{{\textbf{p}}^{{t_0}}}}$ and ${{{\boldsymbol{\chi}}^{{t_0}}}}$ are obtained by solving (P2) with given $\textbf{e}_{\boldsymbol{\theta}} ^{{t_0} - 1}$ according to \textbf{Algorithm 1}.

Similarly, with given ${{{\bf{p}}^{{t_0}}}}$ and ${{{\boldsymbol{\chi}}^{{t_0}}}}$, the following inequality holds
 \begin{align}\label{budengshi2}
 {R_{\rm P2}}\left( {{{\textbf{p}}^{{t_0}}},{{\boldsymbol{\chi}}^{{t_0}}},\textbf{e}_{\boldsymbol{\theta}} ^{{t_0} - 1}} \right) \le {R_{\rm P2}}\left( {{{\textbf{p}}^{{t_0}}},{{\boldsymbol{\chi}}^{{t_0}}},\textbf{e}_{\boldsymbol{\theta}} ^{{t_0}}} \right) = { R_{\rm P1}}\left( {{{\textbf{p}}^{{t_0}}},{{\boldsymbol{\Theta }}^{{t_0}}}} \right).
 \end{align}

From ~\eqref{budengshi1} and ~\eqref{budengshi2}, we have
\begin{align}\label{budengshi3}
{R_{\rm P1}}\left( {{{\textbf{p}}^{{t_0} - 1}},{{\boldsymbol{\Theta }}^{{t_0} - 1}}} \right) \le {R_{\rm P1}}\left( {{{\textbf{p}}^{{t_0}}},{{\boldsymbol{\Theta }}^{{t_0}}}} \right).
 \end{align}

 The inequality in~\eqref{budengshi3} indicates that the objective value of (P1) is monotonically non-decreasing after each iteration. On the other hand, the system throughput is upper bounded. Therefore, the proposed algorithm is guaranteed to converge.
 \section{Numerical Results}
\begin{figure}[!t]
\centering
\includegraphics[scale=0.6]{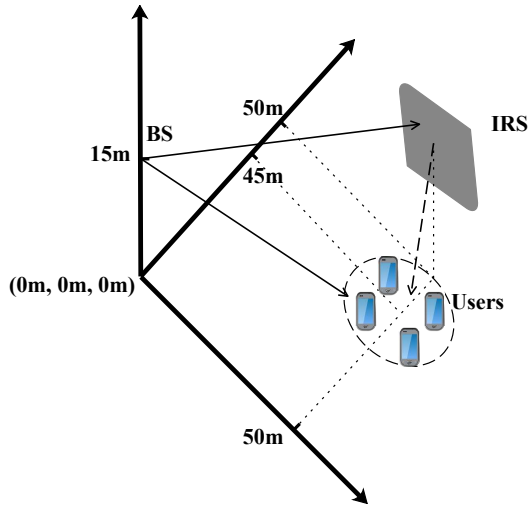}
\caption{Simulation setup for the IRS-NOMA system.}
\label{Simulation_setup}
\end{figure}

In this section, the performances of the proposed algorithms for IRS-NOMA system are evaluated through numerical simulations. The considered downlink IRS-NOMA system scenario is illustrated in Fig. \ref{Simulation_setup}. The BS and IRS are located at coordinates (0m, 0m, 15m) and (50m, 50m, 15m), respectively. The mobile users are randomly and uniformly placed in a circle centered at (50m, 45m, 0m) with radius 5m. The path loss model is ${{\cal P}_{loss}}\left( d \right) = {10^{ - 3}}{\left( d \right)^{ - \alpha }}$, where \emph{d} is the link distance, $\alpha$ is the path loss exponent. The path loss exponents for BS-User link, BS-IRS link and IRS-User link are 3, 2.2 and 2.5, respectively~\cite{fu2019reconfigurable,mu2019exploiting}. The Rician factor $\kappa$ is set to 3dB. The minimum capacity requirement is given by $R_{\rm min}=\rm 0.01bit/s/Hz$, the bandwidth of each channel is 15kHz and the noise power is ${\sigma ^2} =  - 80 \rm dBm$. Without loss of generality, let the maximum number of users allocated to each channel be equal to ${K_e}$, i.e., ${K_n} = {K_{\rm e}}$.

As a benchmark, we also propose a two-step resource allocation algorithm for IRS-OMA systems. The resource allocation problem for IRS-OMA systems is decomposed into two subproblems, including the channel assignment problem, and the joint power allocation and reflection coefficients design problem. The proposed algorithms to solve the above two subproblems are summarized in \textbf{Table~\ref{Table:Referred Algorithms}}.  We refer to the proposed algorithm for IRS-NOMA systems as ThreeStep-IRS-NOMA and refer to the proposed algorithm for IRS-OMA systems as TwoStep-IRS-OMA, respectively. All the referred benchmark algorithms are summarized in \textbf{Table~\ref{Table:Referred Algorithms}}. For example, we refer to the Exhaust-IRS-NOMA algorithm, which solves the channel assignment and decoding order optimization problem by exhaustive search method.

 \begin{table}[htpb]
 \caption{Referred Algorithms}
 \label{Table:Referred Algorithms}
\begin{center}
\begin{tabular} {|c|p{60pt}<{\centering}|p{60pt}<{\centering}|p{60pt}<{\centering}|p{65pt}<{\centering}|p{70pt}<{\centering}|}
\hline
Algorithm &Channel assignment&Decoding order  &Power allocation   &Reflection coefficients  &Communication systems\\
 \hline
Exhaust-IRS-NOMA &Exhaustive search	    &Exhaustive search	 &Algorithm 1	           &Algorithm 3	         &IRS-NOMA\\
 \hline
ThreeStep-IRS-NOMA	 &Algorithm 4	        &Algorithm 5	     &Algorithm 1	           &Algorithm 3	          &IRS-NOMA\\
 \hline
Random-IRS-NOMA &Algorithm 4            &Random selection    &Algorithm 1	           &Algorithm 3	          &IRS-NOMA\\
 \hline
Exhaust-IRS-OMA 	 &Exhaustive search	    &-------- 	         &Water-filling            &Algorithm 3             &IRS-OMA\\
 \hline
TwoStep-IRS-OMA	 &Algorithm 4	        &--------	         &Water-filling            &Algorithm 3             &IRS-OMA\\
  \hline
 NOMA-noIRS	     &Algorithm 4	        &Ordering of channel gains	         &Algorithm 1	           &--------	      &NOMA without IRS\\
  \hline
 OMA-noIRS	     &Algorithm 4	        &--------	         &Water-filling            &--------	         &OMA without IRS\\
 \hline
\end{tabular}
\end{center}
\end{table}
\subsection{Performance of the Proposed Channel Assignment Algorithm}
 \begin{figure}[!t]
\centering
\includegraphics[scale=0.6]{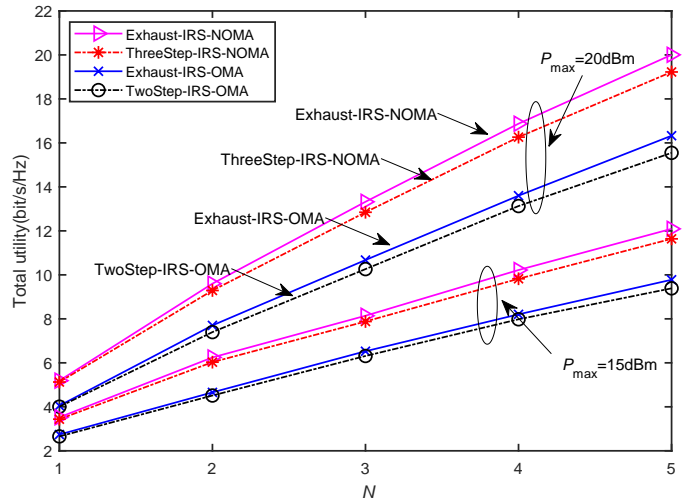}
\caption{Total utility versus the number of channels, ${K_e} = 3$, $M = 80$.}
\label{fig:utility_vs_N}
\end{figure}
We start by presenting the performance of the proposed channel assignment algorithm based on many-to-one matching. Fig. \ref{fig:utility_vs_N} plots the total utility of all channels versus the number of channels \emph{N}. To show the effectiveness of the proposed channel assignment algorithm, we compare it with the exhaustive search based algorithms, i.e., Exhaust-IRS-NOMA and Exhaust-IRS-OMA algorithms. As it can been seen in Fig. \ref{fig:utility_vs_N}, as the number of available channels increases, the total utility of all algorithms increases. The reason is that the users can benefit from channel diversity in the wireless communication environment. In addition, the exhaustive search based algorithms always outperform the non-exhaustive based algorithms. However, with its low complexity, our proposed algorithm can achieve very close performance to that achieved by the exhaustive search based algorithm. Specially, when $N = 4$ and $\emph{P}_{\rm max}={\rm 15dBm}$, the ThreeStep-IRS-NOMA and TwoStep-IRS-OMA achieve around $96\%$ and $97.3\%$ of the utility achieved by the Exhaust-IRS-NOMA and Exhaust-IRS-OMA, respectively.
\subsection{Convergence Performance of the Proposed Algorithm}
\begin{figure}[!t]
\centering
\includegraphics[scale=0.6]{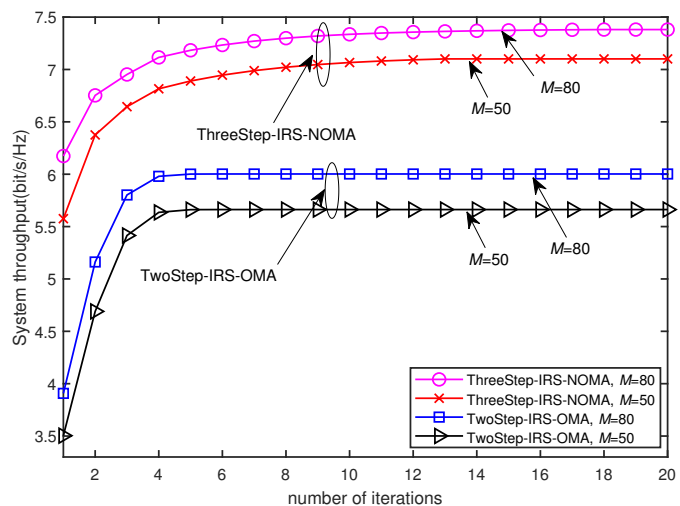}
\caption{Convergence of the proposed algorithms, $N = 2$, ${K_{\rm{e}}} = 3$, ${P_{\max }} = 15{\rm dBm}$.}
\label{fig:Convergence}
\end{figure}
The convergence of the ThreeStep-IRS-NOMA and TwoStep-IRS-OMA versus iteration number are depicted in Fig. \ref{fig:Convergence}, which illustrates that both algorithms can converge within a small number of iterations. Specifically, ThreeStep-IRS-NOMA and TwoStep-IRS-OMA converge in less than 20 and 10 iterations, respectively. Furthermore, the TwoStep-IRS-OMA algorithm converges faster than the ThreeStep-IRS-NOMA algorithm. This is because TwoStep-IRS-OMA has a lower computational complexity than the ThreeStep-IRS-NOMA algorithm. In addition, the number of iterations for the convergence of the two algorithms increases with the number of reflection coefficients, because more variables have to be optimized. For example, when $\emph{M}=80$ and $\emph{M}=50$, the ThreeStep-IRS-NOMA needs 16 and 13 iterations to converge, respectively.
\subsection{Performance Comparison}
Here, we compare the proposed algorithm with benchmark algorithms in \textbf{Table I}.
\subsubsection{System throughput versus the number of passive reflecting elements}
\begin{figure}[!t]
\centering
\includegraphics[scale=0.6]{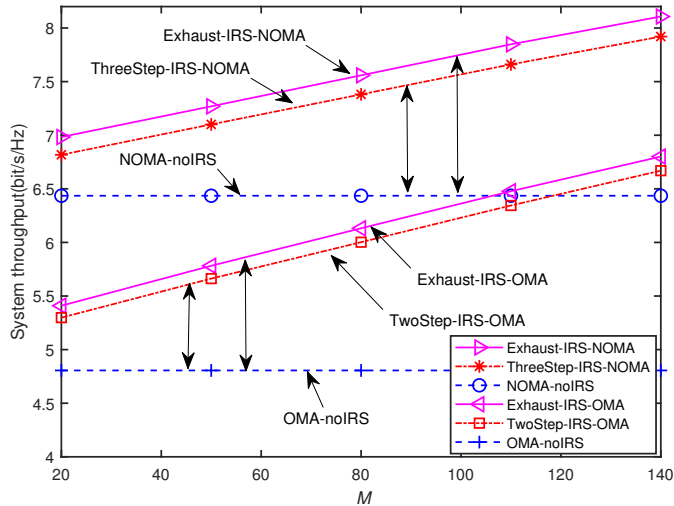}
\caption{System throughput verses the number of reflecting elements, $N = 2$, ${K_{\rm{e}}} = 3$, ${P_{\max }} = 15 {\rm dBm}$.}
\label{fig:Sum rate versus M}
\end{figure}

 In Fig. \ref{fig:Sum rate versus M}, we compare the system throughput performance of various algorithms versus the number of reflecting elements \emph{M}. As it can be observed, the system throughput achieved by the IRS aided algorithms increases with \emph{M}, and significantly outperforms the other algorithms without IRS. This indicates that with more reflecting elements, the resource allocation for IRS assisted systems becomes more flexible and thus achieves higher gains. It can also be observed that the IRS-NOMA system outperforms the IRS-OMA system. Furthermore, IRS aided algorithms achieve significant throughput gains with large \emph{M}. The reason is that more IRS passive reflecting elements can reflect more power of the signals received from the BS which leads to more power gain. In particular, when $M = 20$, the ThreeStep-IRS-NOMA and TwoStep-IRS-OMA algorithms achieve about 0.38bit/s/Hz and 0.49bit/s/Hz performance gain over NOMA-noIRS and OMA-noIRS algorithms, respectively. However, for large \emph{M}, i.e., $M=140$, the performance gains of the two algorithms increase up to 1.49bit/s/Hz and 1.86bit/s/Hz, respectively. In addition, the ThreeStep-IRS-NOMA and TwoStep-IRS-OMA algorithms perform very close to the corresponding exhaustive search based algorithms. For example, when $\emph{M}=80$, the proposed ThreeStep-IRS-NOMA algorithm achieves around $97.6\%$ of the system throughput achieved by Exhaust-IRS-NOMA algorithm.
 \subsubsection{System throughput versus the transmit power budget}
 \begin{figure}[!t]
\centering
\includegraphics[scale=0.6]{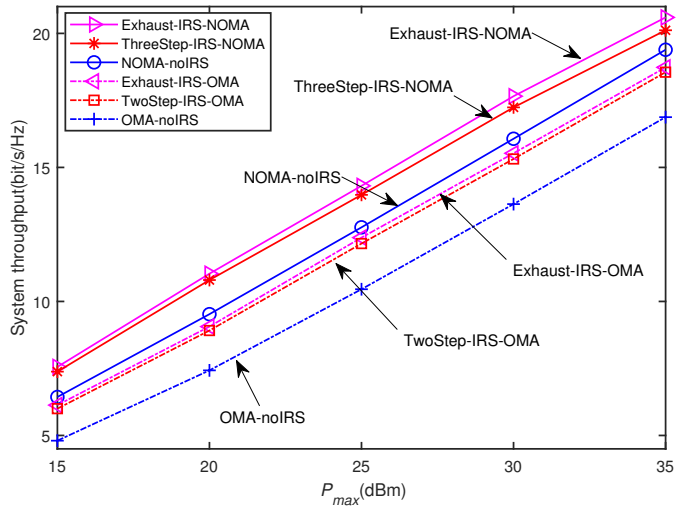}
\caption{System throughput verses the transmit power budget, $N = 2$, ${K_{\rm{e}}} = 3$, $M = 80$.}
\label{fig:rate_vs_power}
\end{figure}
Fig. \ref{fig:rate_vs_power} illustrates the impact of the total transmit power budget ${P_{\max }}$ on the system throughput when $\emph{N}=2$, $\emph{K}=3$ and $\emph{M}=80$. It is observed that the system throughput of all considered algorithms increases with the increase of ${P_{\max }}$. The IRS-assisted algorithms significantly outperform the algorithms without the IRS, which confirms the advantages of introducing the IRS. In comparison with OMA based algorithms, the NOMA based algorithms, including NOMA-noIRS, ThreeStep-IRS-NOMA and Exhaust-IRS-NOMA yield a significant performance gain, because NOMA allows the users to access the same channel and hence the performance can be improved. Furthermore, the ThreeStep-IRS-NOMA and TwoStep-IRS-OMA algorithms achieve near-optimal performance as the Exhaust-IRS-NOMA and Exhaust-IRS-OMA, respectively. This result can also been observed in Fig. \ref{fig:utility_vs_N} and Fig. \ref{fig:Sum rate versus M}.
\subsubsection{Impact of decoding order}
\begin{figure}[!t]
\centering
\includegraphics[scale=0.6]{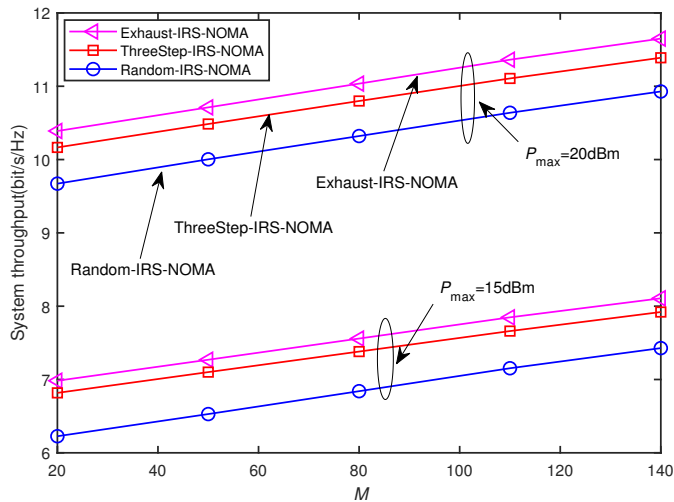}
\caption{System throughput verses the number of passive reflecting elements, $N = 2$, ${K_{\rm{e}}} = 3$.}
\label{fig:rate_vs_decoding}
\end{figure}
Now, we evaluate the impact of the decoding order on the system throughput performance. Two algorithms are compared with our proposed low-complexity decoding order optimization algorithm. The first one is the Exhaust-IRS-NOMA algorithm, which finds the optimal decoding order via exhaustive search. The second one is the Random-IRS-NOMA algorithm, which randomly selects the decoding order. As expected, we can see from Fig. \ref{fig:rate_vs_decoding} that, the Exhaust-IRS-NOMA algorithm outperforms the other two algorithms over the entire range of the number of passive reflecting elements \emph{M}, which demonstrates the importance of finding the optimal decoding order. However, the Exhaust-IRS-NOMA algorithm needs to search ${{K_n}!}$ possible decoding orders for each channel assignment, which is very complex. The proposed ThreeStep-IRS-NOMA algorithm can achieve a similar performance as the Exhaust-IRS-NOMA algorithm with low complexity and outperforms the Random-IRS-NOMA algorithm, which illustrates the effectiveness of the proposed algorithm in determining the decoding order.
\subsubsection{Impact of the IRS location}
\begin{figure}[!t]
\centering
\includegraphics[scale=0.6]{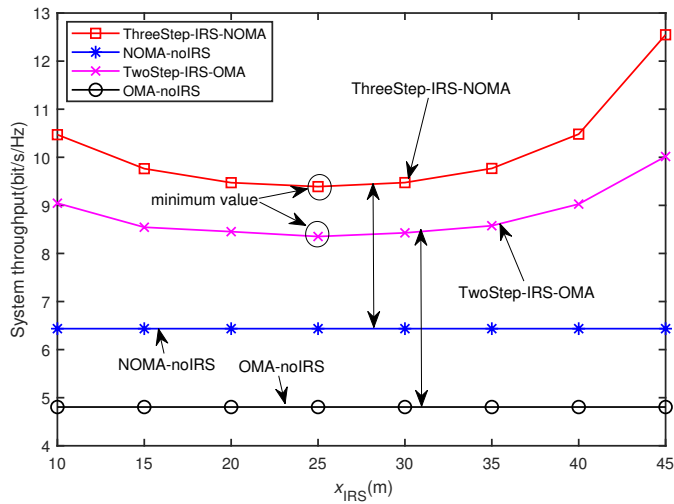}
\caption{System throughput versus the location of IRS coordinate, $N = 2$, ${K_{\rm{e}}} = 3$, ${P_{\max }} = 15{\rm dBm}$, $M = 80$.}
\label{fig:rate_vs_dis}
\end{figure}
In this part, we study the impact of the IRS location. Without loss of generality, we set the coordinates of the BS location and the IRS location as (0m, 0m, 0m) and ($\emph{x}_{\rm IRS}$, 0m, 0m), respectively. The distance between the BS and the user central point is $d_{\rm User}^{\rm BS} = 50\rm m$. The path loss exponents of BS-IRS link and IRS-User link are both equal to 2.5. The location of IRS is moved from ${x_{\rm IRS}} =10 \rm m $ to ${x_{\rm IRS}} =45\rm m$. As shown in Fig. \ref{fig:rate_vs_dis}, the system throughput of ThreeStep-IRS-NOMA and TwoStep-IRS-OMA first decrease and then increase after achieving their minimum system throughput at ${x_{\rm IRS}} =25\rm m$. To simplify the analysis, we ignore the small-scale fading effects. Then, the large-scale channel gain of BS-IRS-User link can be simply approximated as
 \begin{align}\label{approximated pathloss}
{\cal P} = {10^{ - 6}}{\left( {d_{\rm IRS}^{\rm BS}d_{\rm User}^{\rm IRS}} \right)^{ - 2.5}}{\rm{ + }}{10^{ - 3}}{\left( {d_{\rm User}^{\rm BS}} \right)^{ - 3}},
 \end{align}
 where $d_{\rm IRS}^{\rm BS} + d_{\rm User}^{\rm IRS} = d_{\rm User}^{\rm BS}$. When $d_{\rm IRS}^{\rm BS}{\rm{ = }}d_{\rm User}^{\rm IRS}{\rm{ = }}{{d_{\rm User}^{\rm BS}} \mathord{\left/  {\vphantom {{d_{\rm User}^{\rm BS}} 2}} \right.  \kern-\nulldelimiterspace} 2}$, the combined channel gain achieves its minimum value at the middle point, which explains the simulation results in Fig. \ref{fig:rate_vs_dis}. The system throughput gain of ThreeStep-IRS-NOMA over NOMA-noIRS is 4.1bit/s/Hz at ${x_{\rm IRS}} =10\rm m$, while the performance gain increases to 6.1bit/s/Hz at  ${x_{\rm IRS}} =45\rm m$. Because there is a strong IRS-user link when the IRS is near the user side. Therefore, the system performance can be significantly enhanced by carefully choosing the location of the IRS.
 \section{Conclusions}
The resource allocation problem of the downlink transmissions in  the IRS-NOMA system has been investigated in this paper. The system throughput maximization problem was formulated by jointly optimizing the channel assignment, decoding order, power allocation and reflection coefficients. In particular, the original problem was divided into three subproblems and solved sequentially where: 1) a low-complexity channel assignment algorithm based on matching theory has been proposed, which can achieve a near-optimal performance as the exhaust search algorithm; 2) we proposed a low-complexity decoding order optimization algorithm, which can achieve a comparable performance to the exhaustive decoding order search method; 3) by invoking alternating optimization and successive convex approximation, we proposed an efficient power allocation and reflection coefficients design algorithm. Simulation results showed that the proposed algorithms can improve the system throughput of the novel IRS-NOMA system. Our results confirm that introducing IRS, IRS-NOMA systems outperform traditional NOMA systems.
\section*{Appendix~A: Proof of Theorem 1} \label{Theorem NP}
\renewcommand{\theequation}{A.\arabic{equation}}
\setcounter{equation}{0}
It is well known, three-dimensional matching is NP-hard problem~\cite{cui2017optimal}. If we prove that a special three-dimensional matching case of problem (P1) is NP-hard, then the original problem (P1) is also NP-hard problem.

For the given power allocation and reflection coefficients, let the maximum number of users assigned to each channel be ${K_n} = 2$ and divide the user set ${\cal K}$ into two disjoint sets ${{\cal K}_1^{\rm sub}}$ and ${{\cal K}_2^{\rm sub}}$, where the size $\left| {{\cal K}_1^{\rm sub}} \right| = \left| {{\cal K}_2^{\rm sub}} \right| = {K \mathord{\left/ {\vphantom {K 2}} \right. \kern-\nulldelimiterspace} 2}$, $\left| {{\cal K}_1^{\rm sub}} \right| \cup \left| {{\cal K}_2^{\rm sub}} \right| = {\cal K}$ and $\left| {{\cal K}_1^{\rm sub}} \right| \cap \left| {{\cal K}_2^{\rm sub}} \right| = \emptyset$. Channel \emph{n} chooses two users ${k_{n,1}} \in \mathcal{K}_1^{{\rm{sub}}}$ and ${k_{n,2}} \in \mathcal{K}_2^{{\rm{sub}}}$. Define ${\cal D} = \left\{ {{{\cal D}_1},{{\cal D}_2}, \cdots ,{{\cal D}_N}} \right\}$ as a subset of ${\cal N} \times {\cal K}_1^{\rm sub} \times {\cal K}_2^{\rm sub}$, where the triplet ${{\cal D}_n} = \left( {n,{k_{n,1}},{k_{n,2}}} \right)$. According to~\cite{cui2017optimal}, there exist a subset $\widetilde {\cal D} \subseteq {\cal D}$ satisfying the following conditions: 1) $\left| {\widetilde {\cal D}} \right| = \min \left\{ {N,\frac{K}{2}} \right\}$; 2) for any two distinct triplets $\left( {n,{k_{n,1}},{k_{n,2}}} \right) \in {\cal D}$ and $\left( {\widetilde n,{k_{\widetilde n,1}},{k_{\widetilde n,2}}} \right) \in \widetilde {\cal D}$, we have $n \ne \widetilde n$, ${k_{n,1}} \ne {k_{\widetilde n,1}}$ and ${k_{n,2}} \ne {k_{\widetilde n,2}}$. Then, $\widetilde {\cal D}$ is a three-dimensional matching and the constructed special case of problem (P1) is NP-hard. Therefore, (P1) is also NP-hard.

\section*{Appendix~B: Proof of Theorem 2} \label{Theorem convergence of matching}
\renewcommand{\theequation}{B.\arabic{equation}}
\setcounter{equation}{0}
A matching is named two-sided stable if there does not exist a swap-blocking pair. In \textbf{Algorithm 4}, the convergence mainly depends on the second swapping process. According to the swap block pair definition, after each swap operation between user \emph{k} in channel \emph{n} and user $\widetilde k$ in channel $\widetilde n$, we have $\forall \omega  \in \left\{ {n,\widetilde n} \right\},{U_\omega }\left( {\Upsilon _k^{\widetilde k}} \right) \ge {U_\omega }\left( \Upsilon  \right)$ and $\exists \omega  \in \left\{ {n,\widetilde n} \right\},{U_\omega }\left( {\Upsilon _k^{\widetilde k}} \right) > {U_\omega }\left( \Upsilon  \right)$, which means that at least one channel utility will increase. Furthermore, the total utility of all channels satisfies the following inequality: $\sum\limits_{n = 1}^N {{U_n}\left( {\Upsilon _k^{\widetilde k}} \right)}  > \sum\limits_{n = 1}^N {{U_n}\left( \Upsilon  \right)} $. Since the numbers of users and channels are limited, and the total utility is upper bounded due to the transmit power budget, the number of swap block pairs is  limited. Therefore, when there is no swap operation, \textbf{Algorithm 4} will converge.

\vspace{-0.5cm}
\bibliographystyle{IEEEtran}
\bibliography{zjkbib}

\begin{thebibliography}{10}
\providecommand{\url}[1]{#1}
\csname url@samestyle\endcsname
\providecommand{\newblock}{\relax}
\providecommand{\bibinfo}[2]{#2}
\providecommand{\BIBentrySTDinterwordspacing}{\spaceskip=0pt\relax}
\providecommand{\BIBentryALTinterwordstretchfactor}{4}
\providecommand{\BIBentryALTinterwordspacing}{\spaceskip=\fontdimen2\font plus
\BIBentryALTinterwordstretchfactor\fontdimen3\font minus
  \fontdimen4\font\relax}
\providecommand{\BIBforeignlanguage}[2]{{%
\expandafter\ifx\csname l@#1\endcsname\relax
\typeout{** WARNING: IEEEtran.bst: No hyphenation pattern has been}%
\typeout{** loaded for the language `#1'. Using the pattern for}%
\typeout{** the default language instead.}%
\else
\language=\csname l@#1\endcsname
\fi
#2}}
\providecommand{\BIBdecl}{\relax}
\BIBdecl

\bibitem{Zuo2020}
J.~Zuo, Y.~Liu, Z.~Qin, and C.~Shen, ``The application of intelligent
  reflecting surface in downlink {NOMA} systems,'' \emph{{\rm in} Proc. IEEE
  ICC Workshop}, 2020.

\bibitem{liu2018non}
Y.~Liu, Z.~Qin, M.~Elkashlan, Z.~Ding, A.~Nallanathan, and L.~Hanzo,
  ``Non-orthogonal multiple access for {5G} and beyond,'' \emph{Proc. IEEE},
  vol. 105, no.~12, pp. 2347--2381, 2017.

\bibitem{3GPPStudy}
3rd Generation Partnership Project~(3GPP), ``Study on downlink multiuser
  superposition transmission for {LTE},'' 2015.

\bibitem{dai2018survey}
L.~Dai, B.~Wang, Z.~Ding, Z.~Wang, S.~Chen, and L.~Hanzo, ``A survey of
  non-orthogonal multiple access for {5G},'' \emph{IEEE Commun Surv. Tut.},
  vol.~20, no.~3, pp. 2294--2323, 2018.

\bibitem{qin2018user}
Z.~Qin, X.~Yue, Y.~Liu, Z.~Ding, and A.~Nallanathan, ``User association and
  resource allocation in unified {NOMA} enabled heterogeneous ultra dense
  networks,'' \emph{IEEE Commun. Mag.}, vol.~56, no.~6, pp. 86--92, 2018.

\bibitem{ding2017application}
Z.~Ding, Y.~Liu, J.~Choi, Q.~Sun, M.~Elkashlan, I.~Chih-Lin, and H.~V. Poor,
  ``Application of non-orthogonal multiple access in {LTE} and {5G} networks,''
  \emph{IEEE Commun. Mag.}, vol.~55, no.~2, pp. 185--191, 2017.

\bibitem{liu2017non}
Y.~Liu, Z.~Qin, M.~Elkashlan, A.~Nallanathan, and J.~A. McCann,
  ``Non-orthogonal multiple access in large-scale heterogeneous networks,''
  \emph{{IEEE} J. Sel. Areas Commun.}, vol.~35, no.~12, pp. 2667--2680, 2017.

\bibitem{gong2019towards}
S.~Gong, X.~Lu, D.~T. Hoang, D.~Niyato, L.~Shu, D.~I. Kim, and Y.-C. Liang,
  ``Towards smart radio environment for wireless communications via intelligent
  reflecting surfaces: a comprehensive survey,'' \emph{arXiv preprint
  arXiv:1912.07794}, 2019.

\bibitem{zhao2019survey}
J.~Zhao, ``A survey of intelligent reflecting surfaces {(IRSs)}: towards {6G}
  wireless communication networks with massive {MIMO} 2.0,'' 2019.

\bibitem{tang2019wireless}
W.~Tang, M.~Z. Chen, X.~Chen, J.~Y. Dai, Y.~Han, M.~Di~Renzo, Y.~Zeng, S.~Jin,
  Q.~Cheng, and T.~J. Cui, ``Wireless communications with reconfigurable
  intelligent surface: path loss modeling and experimental measurement,''
  \emph{arXiv preprint arXiv:1911.05326}, 2019.

\bibitem{jung2018performance}
M.~Jung, W.~Saad, Y.~Jang, G.~Kong, and S.~Choi, ``Performance analysis of
  large intelligence surfaces {(LISs)}: asymptotic data rate and channel
  hardening effects,'' \emph{arXiv preprint arXiv:1810.05667}, 2018.

\bibitem{di2016sub}
B.~Di, L.~Song, and Y.~Li, ``Sub-channel assignment, power allocation, and user
  scheduling for non-orthogonal multiple access networks,'' \emph{{IEEE} Trans.
  Wireless Commun.}, vol.~15, no.~11, pp. 7686--7698, 2016.

\bibitem{zhao2017spectrum}
J.~Zhao, Y.~Liu, K.~K. Chai, A.~Nallanathan, Y.~Chen, and Z.~Han, ``Spectrum
  allocation and power control for non-orthogonal multiple access in
  {HetNets},'' \emph{{IEEE} Trans. Wireless Commun.}, vol.~16, no.~9, pp.
  5825--5837, 2017.

\bibitem{liu2018super}
G.~Liu, R.~Wang, H.~Zhang, W.~Kang, T.~A. Tsiftsis, and V.~C. Leung,
  ``Super-modular game-based user scheduling and power allocation for
  energy-efficient {NOMA} network,'' \emph{{IEEE} Trans. Wireless Commun.},
  vol.~17, no.~6, pp. 3877--3888, 2018.

\bibitem{shi2019energy}
J.~Shi, W.~Yu, Q.~Ni, W.~Liang, Z.~Li, and P.~Xiao, ``Energy efficient resource
  allocation in hybrid non-orthogonal multiple access systems,'' \emph{{IEEE}
  Trans. Commun.}, vol.~67, no.~5, pp. 3496--3511, 2019.

\bibitem{fang2018joint}
F.~Fang, J.~Cheng, and Z.~Ding, ``Joint energy efficient subchannel and power
  optimization for a downlink {NOMA} heterogeneous network,'' \emph{{IEEE}
  Trans. Veh. Commun.}, vol.~68, no.~2, pp. 1351--1364, 2018.

\bibitem{sun2017optimal}
Y.~Sun, D.~W.~K. Ng, Z.~Ding, and R.~Schober, ``Optimal joint power and
  subcarrier allocation for full-duplex multicarrier non-orthogonal multiple
  access systems,'' \emph{{IEEE} Trans. Commun.}, vol.~65, no.~3, pp.
  1077--1091, 2017.

\bibitem{sun2018robust}
Y.~Sun, D.~W.~K. Ng, J.~Zhu, and R.~Schober, ``Robust and secure resource
  allocation for full-duplex {MISO} multicarrier {NOMA} systems,'' \emph{{IEEE}
  Trans. Commun.}, vol.~66, no.~9, pp. 4119--4137, 2018.

\bibitem{ning2019intelligent}
B.~Ning, Z.~Chen, W.~Chen, and J.~Fang, ``Intelligent reflecting surface design
  for {MIMO} system by maximizing sum-path-gains,'' \emph{arXiv preprint
  arXiv:1909.07282}, 2019.

\bibitem{zhang2019capacity}
S.~Zhang and R.~Zhang, ``Capacity characterization for intelligent reflecting
  surface aided mimo communication,'' \emph{arXiv preprint arXiv:1910.01573},
  2019.

\bibitem{chu2019intelligent}
Z.~Chu, W.~Hao, P.~Xiao, and J.~Shi, ``Intelligent reflecting surface aided
  multi-antenna secure transmission,'' \emph{{IEEE} Wireless Commun. Lett.},
  2019.

\bibitem{yang2019intelligent}
Y.~Yang, B.~Zheng, S.~Zhang, and R.~Zhang, ``Intelligent reflecting surface
  meets {OFDM}: protocol design and rate maximization,'' \emph{arXiv preprint
  arXiv:1906.09956}, 2019.

\bibitem{li2019reconfigurable}
S.~Li, B.~Duo, X.~Yuan, Y.-C. Liang, M.~Di~Renzo \emph{et~al.},
  ``Reconfigurable intelligent surface assisted {UAV} communication: joint
  trajectory design and passive beamforming,'' \emph{arXiv preprint
  arXiv:1908.04082}, 2019.

\bibitem{wu2019joint}
Q.~Wu and R.~Zhang, ``Joint active and passive beamforming optimization for
  intelligent reflecting surface assisted {SWIPT} under {QoS} constraints,''
  \emph{arXiv preprint arXiv:1910.06220}, 2019.

\bibitem{yang2019intelligentNOMA}
G.~Yang, X.~Xu, and Y.-C. Liang, ``Intelligent reflecting surface assisted
  non-orthogonal multiple access,'' \emph{arXiv preprint arXiv:1907.03133},
  2019.

\bibitem{ding2019simple}
Z.~Ding and H.~V. Poor, ``A simple design of {IRS-NOMA} transmission,''
  \emph{arXiv preprint arXiv:1907.09918}, 2019.

\bibitem{li2019joint}
Y.~Li, M.~Jiang, Q.~Zhang, and J.~Qin, ``Joint beamforming design in
  multi-cluster {MISO NOMA} intelligent reflecting surface-aided downlink
  communication networks,'' \emph{arXiv preprint arXiv:1909.06972}, 2019.

\bibitem{mu2019exploiting}
X.~Mu, Y.~Liu, L.~Guo, J.~Lin, and N.~Al-Dhahir, ``Exploiting intelligent
  reflecting surfaces in multi-antenna aided {NOMA} systems,'' \emph{arXiv
  preprint arXiv:1910.13636}, 2019.

\bibitem{zhu2019power}
J.~Zhu, Y.~Huang, J.~Wang, K.~Navaie, and Z.~Ding, ``Power efficient
  {IRS}-assisted {NOMA},'' \emph{arXiv preprint arXiv:1912.11768}, 2019.

\bibitem{fu2019reconfigurable}
M.~Fu, Y.~Zhou, and Y.~Shi, ``Reconfigurable intelligent surface empowered
  downlink non-orthogonal multiple access,'' \emph{arXiv preprint
  arXiv:1910.07361}, 2019.

\bibitem{liu2016fairness}
Y.~Liu, M.~Elkashlan, Z.~Ding, and G.~K. Karagiannidis, ``Fairness of user
  clustering in {MIMO} non-orthogonal multiple access systems,'' \emph{{IEEE}
  Commun. Lett.}, vol.~20, no.~7, pp. 1465--1468, 2016.

\bibitem{cui2017optimal}
J.~Cui, Y.~Liu, Z.~Ding, P.~Fan, and A.~Nallanathan, ``Optimal user scheduling
  and power allocation for millimeter wave {NOMA} systems,'' \emph{{IEEE}
  Trans. Wireless Commun.}, vol.~17, no.~3, pp. 1502--1517, 2017.

\bibitem{tran2012fast}
L.-N. Tran, M.~F. Hanif, A.~Tolli, and M.~Juntti, ``Fast converging algorithm
  for weighted sum rate maximization in multicell {MISO} downlink,''
  \emph{{IEEE} Signal Process. Lett.}, vol.~19, no.~12, pp. 872--875, 2012.

\bibitem{grant2014cvx}
M.~Grant and S.~Boyd, ``{CVX}: Matlab software for disciplined convex
  programming, version 2.1,'' 2014.

\bibitem{liu2017enhancing}
Y.~Liu, Z.~Qin, M.~Elkashlan, Y.~Gao, and L.~Hanzo, ``Enhancing the physical
  layer security of non-orthogonal multiple access in large-scale networks,''
  \emph{{IEEE} Trans. Wireless Commun.}, vol.~16, no.~3, pp. 1656--1672, 2017.

\end{thebibliography}

\end{document}